\documentclass[a4paper, amsfonts, amssymb, amsmath, reprint, showkeys, showpacs, nofootinbib, twoside, superscriptaddress,aps,prd]{revtex4-1}
\usepackage[english]{babel}
\usepackage[utf8]{inputenc}
\usepackage{comment}
\usepackage[colorinlistoftodos, color=green!40, prependcaption]{todonotes}
\usepackage{amsthm}
\usepackage{mathtools}
\usepackage{physics}
\usepackage{xcolor}
\usepackage{graphicx}
\usepackage{adjustbox}
\usepackage{placeins}
\usepackage[T1]{fontenc}
\usepackage{lipsum}
\usepackage{csquotes}
\usepackage{enumerate}

\usepackage{ulem}

\DeclareMathOperator\arctanh{atanh}

\begin{document}
\title{Strong deflection limit analysis of black hole lensing in inhomogeneous plasma}

\author{Fabiano Feleppa}
    \email[]{ffeleppa@unisa.it}
     \affiliation{Dipartimento di Fisica “E.R. Caianiello”, Università di Salerno, Via Giovanni Paolo II 132, I-84084 Fisciano, Italy}
    \affiliation{Istituto Nazionale di Fisica Nucleare, Sezione di Napoli, Via Cintia, 80126, Napoli, Italy}

    \author{Valerio Bozza}
    \email[]{vbozza@unisa.it}
    \affiliation{Dipartimento di Fisica “E.R. Caianiello”, Università di Salerno, Via Giovanni Paolo II 132, I-84084 Fisciano, Italy}
    \affiliation{Istituto Nazionale di Fisica Nucleare, Sezione di Napoli, Via Cintia, 80126, Napoli, Italy}

    \author{Oleg Yu.\ Tsupko}
    \email[]{tsupkooleg@gmail.com}
    \affiliation{ZARM, University of Bremen, 28359 Bremen, Germany}

\begin{abstract}
This paper investigates gravitational lensing effects in the presence of plasma in the strong deflection limit, which corresponds to light rays circling around a compact object and forming higher-order images. While previous studies of this case have predominantly focused on the deflection of light in a vacuum or in the presence of a homogeneous plasma, this work introduces an analytical treatment for the influence of a non-uniform plasma. After recalling the exact expression for the deflection angle of photons in a static, asymptotically flat and spherically symmetric spacetime filled with cold non-magnetized plasma, a strong deflection limit analysis is presented. Particular attention is then given to the case of a Schwarzschild spacetime, where the deflection angle of photons for different density profiles of plasma is obtained. Moreover, perturbative results for an arbitrary power-law radial density profile are also presented. These formulas are then applied to the calculation of the positions and magnifications of higher-order images, concluding that the presence of a non-uniform plasma reduces both their angular size and their magnifications, at least within the range of the power-law indices considered. These findings contribute to the understanding of gravitational lensing in the presence of plasma, offering a versatile framework applicable to various asymptotically flat and spherically symmetric spacetimes.
\end{abstract}


\maketitle

\section{Introduction}
\label{sec:intro}

Gravitational deflection, one of the earliest phenomena explored within the general theory of relativity, was initially observed in the bending of light around the Sun. Subsequently, it was identified, e.g., in the lensing of quasars by foreground galaxies \cite{Walsh}, in the formation of arcs in galaxy clusters \cite{Soucail}, in galactic microlensing \cite{Paczynski} and other phenomena of gravitational lensing.\ In the weak deflection approximation, the theory has passed all tests with flying colors \cite{Falco1992, Narayan, Wambsganss1998, Schneider2001, Mollerach, Dodelson2017, Meneghetti2021}.

On the other hand, in recent years, the study of light bending by compact objects has gained a significant momentum, due to the groundbreaking observations conducted by the Event Horizon Telescope team \cite{L1,L2,L3,L4,L5,L6, Kocherlakota-2021, L12,L13,L14,L15,L16,L17} (see also Ref.\ \cite{Agol2000} where the authors introduce the idea behind such observations). By leveraging an international network of radio telescopes, the team has provided unprecedented insights into the immediate vicinity of black holes. They captured images that were once thought to be beyond the grasp of observational capabilities. These observations have not only validated the existence of supermassive black holes at the centers of galaxies but have also opened a new era in the study of lensing beyond the weak deflection approximation; this allows one to examine the regions surrounding the black hole event horizon through different techniques, see e.g., Refs.\ \cite{Bambi2019,Tamburini1,Tamburini2,Tamburini3}.

Even if only from a theoretical perspective, the deflection of light due to very compact objects has also been studied for a long time. In 1959, Darwin \cite{Darwin1959} investigated the deflection of light in a Schwarzschild background. In particular, he derived a logarithmic approximation (now referred to as strong deflection limit) for light rays moving near the photon sphere and described the appearance of higher-order images (`ghosts'); see also subsequent studies of Atkinson \cite{Atkinson1965}, Misner, Thorne and Wheeler \cite{Misner1973}, Luminet \cite{Luminet1979} and Ohanian \cite{Ohanian1987}.\ Afterward, using the exact expression for the deflection angle, Virbhadra and Ellis \cite{Ellis2000} numerically calculated the properties of higher-order images (`relativistic images') in the case of a Schwarzschild black hole. In the same year, Frittelli, Kling and Newman \cite{Frittelli2000} obtained solutions to the exact lens equation in the form of integral expressions. The exact gravitational lens equation in spherically symmetric and static spacetimes has also been investigated by Perlick \cite{Perlick2004}. For a detailed discussion on higher-order images and related topics, the reader may refer to Ref.\ \cite{Bozza2010}.

The investigation of higher-order images is highly simplified in the strong deflection limit, which provides an analytical logarithmic approximation for the deflection angle. Calculations of the positions and magnifications of higher-order images were initially performed for a Schwarzschild black hole \cite{Bozza2001} and later generalized to generic spherically symmetric spacetimes \cite{Bozza2002} and rotating black holes \cite{Bozza2003}. From then on, especially regarding the possibility of distinguishing different theories of gravity, numerous studies on gravitational lensing beyond the weak deflection approximation have appeared in the literature (see, e.g., Refs.\ \cite{Claudel2001, Hasse2002, Perlick2004-review, Iyer2007, Keeton2008, Tsupko2008, Majumdar2009, Tarasenko2010, Eiroa2011, Wei2012, Zhang2015, Alhamzawi2016, Tsukamoto2016, Aldi-Bozza-2017, Dai2018, Aratore2021, Kuang2022, Aratore-Bozza-2024}). In particular, higher-order images in the form of photon rings around the black hole shadow have been widely studied (see, e.g., \cite{Gralla2019, Johnson-2020, Gralla2020, Lupsasca2020, Gralla-Lupsasca-2020, Wielgus-2021, Broderick-2022, Ayzenberg-2022, Guerrero-2022, BK-Tsupko-2022, Tsupko-2022, Eichhorn-2023, Broderick-Salehi-2023, Kocherlakota-2024-1, Kocherlakota-2024-2, Aratore-Tsupko-Perlick-2024}).

All the above-mentioned studies on higher-order images and the strong deflection case are based on the assumption that light propagates along lightlike geodesics, without direct influence from matter on the trajectories of rays. However, the presence of plasma in the regions of light propagation changes the ray trajectory due to refraction and dispersion of the medium. In the last decade or so, many works describing different scenarios have appeared in the literature, taking into account such influence, both in the weak field approximation \cite{Tsupko2009, BK-Tsupko-2010, Morozova2013, Er2014, Tsupko2015, Gallo2018, Gallo20191, Gallo20192, Tsupko2020, Sun2023, BK-Tsupko-2023-time-delay} and beyond \cite{Perlick-2000, Tsupko2013, Perlick2015, Rogers2015, Rogers20171, Rogers20172, Perlick2017, Huang2018, Yan2019, Kimpson20191, Kimpson20192, Babar2020, Tsupko2021, Chowdhuri2021, Eiroa2021, Li2022, Bezdekova2022, Zhang2023, Briozzo2023, Gallo2023, Eiroa2023, Bezdekova-2023, Bezdekova-2024, Perlick-Tsupko-2024}; see also earlier works \cite{Synge1960, Muhleman-Johnston-1966, Bicak-1975, Breuer-Ehlers-1980, Breuer-Ehlers-1981a, Breuer-Ehlers-1981b, Bliokh-Minakov-1989, Kulsrud-Loeb-1992, Brod-Blandford-2003a, Brod-Blandford-2003b}. For recent reviews on the topic and on black hole lensing in general, the reader may refer to Refs.\ \cite{Tsupko2017, Cunha2018, Perlick2022}; see also earlier reviews \cite{Perlick2004-review, Bozza2010}. In most of the mentioned literature, both homogeneous and non-homogeneous plasma have been studied; however, when it comes to the analytical calculation of the deflection angle of photons in the strong deflection limit, the only case that has been considered so far is the one of a cold non-magnetized homogeneous plasma \cite{Tsupko2013}.

In this paper, it is our goal to extend the results of Ref.\ \cite{Tsupko2013} to the case of a plasma with an arbitrary radial density profile. At the same time, following the procedure outlined in \cite{Bozza2002}, we will present results that can be easily adapted to any static and spherically symmetric spacetime.

The paper is organized as follows. In Sec.\ \ref{sec:deflangwithpl}, we introduce some notation and in particular recall the expression for the deflection angle of photons in a static, spherically symmetric and asymptotically flat spacetime filled with cold non-magnetized plasma \cite{Tsupko2013}, setting the stage for the rest of the paper. In Sec.\ \ref{sec:sdlpl}, the strong deflection limit procedure introduced in Ref.\ \cite{Bozza2002} is generalized to include matter. Specializing to the case of a Schwarzschild spacetime, in Sec.\ \ref{sec:schwlensing} we will consider different density profiles for the plasma; after reproducing the known result for the deflection angle in the presence of a homogeneous plasma (see Ref.\ \cite{Tsupko2013} for details), we will analyze some cases commonly considered in the literature. We conclude this section showing how semi-analytical results can be obtained for an arbitrary radial power-law density profile. These results are then applied to the calculation of the positions and magnifications of higher-order images in Sec.\ \ref{sec:relimages}. Finally, Sec.\ \ref{sec:concl} is devoted to concluding remarks.

In what follows, we set $G = c = 1$ and work with signature convention $\{-, +, +, +\}$. Moreover, Greek indices sum over the spatial coordinates, while Latin indices run over all four.

\section{Deflection angle of photons in a spacetime filled with cold non-magnetized plasma} 
\label{sec:deflangwithpl}
In his work \cite{Synge1960}, Synge developed a framework for understanding general relativistic geometrical optics within curved spacetime that is filled with an isotropic transparent medium (with negligible self-gravity effects). In this paper, we focus on a particular kind of medium:\ cold, non-magnetized plasma. Furthermore, we are interested in studying lensing by static, spherically symmetric black holes (or any other sufficiently compact object), described by the line element
\begin{align} \label{line-element}
    g_{ik}dx^i dx^k &= g_{00}\left(dx^{0}\right)^2 +  g_{\alpha \beta}dx^\alpha dx^\beta \nonumber \\
    &= -A(r) \, dt^2 + B(r) \, dr^2 + C(r) \, d\Omega^2,
\end{align}
where $d\Omega^2 \coloneqq d\theta^2 + \sin^2 \theta \, d\varphi^2$ defines the round metric on the unit two-sphere. We further assume the spacetime to be asymptotically flat.
In the geometric optics limit, photon trajectories in the presence of both a gravitational field and non-magnetized plasma can be calculated from the variational principle \cite{Synge1960}
\begin{equation}\label{variational principle}
    \delta \int p_i dx^i = 0,
\end{equation}
with $p^i$ being the linear momentum of photons, together with the constraint
\begin{equation}\label{constraint}
    \mathcal{H}(x^i,p_i) = 0,
\end{equation}
where the scalar function $\mathcal{H}(x^i,p_i)$ reads \cite{Perlick-2000, Perlick2015}
\begin{equation}\label{Hamiltonian}
    \mathcal{H}(x^i, p_i)\coloneqq \frac{1}{2}\left(g^{ik}p_i p_k + \omega_e^2 (x^i)\right).
\end{equation}
In the above expression, $\omega_e (x^i)$ represents the plasma frequency, given by
\begin{equation}
    \omega_e^2 (x^i) = \frac{4\pi e^2}{m_e}N(x^i),
\end{equation}
where $e$ and $m_e$ are the electron charge and mass, respectively, and $N(x^i)$ is the electron number density (measured in the frame comoving with the plasma).

Under these assumptions, we consider a photon that moves from infinity toward a spherically symmetric and static central object surrounded by cold plasma and then returns to infinity. Without loss of generality, we assume that the motion occurs in the equatorial plane ($\theta=\pi/2$). An exact expression for the deflection angle of such photon results in \cite{Perlick-2000, Tsupko2013, Perlick2015, Tsupko2021}
\begin{equation}
    \hat{\alpha}(r_0) = 2\int_{r_0}^{\infty}\sqrt{\frac{B(r)}{C(r)}}\left(\frac{h^2 (r)}{h^2 (r_0)} - 1\right)^{-\frac{1}{2}}dr - \pi,  
\end{equation}
where $r_0$ denotes the minimum radial coordinate reached by the photon from the black hole, while the function $h(r)$ is defined as
\begin{equation} \label{h-definition}
    h^2(r) \coloneqq \frac{C(r)}{A(r)}\left(1 - A(r)\frac{\omega_e^2 (r)}{\omega_\infty^2}\right),
\end{equation}
with $\omega_\infty$ being the photon frequency measured by an observer on a $t$-line at infinity. The frequency $\omega(r)$ measured by a static observer is equal to $\omega_\infty/\sqrt{A(r)}$ (see, e.g., \cite{Perlick2015, Bezdekova-2024}). In the above discussion, we also implicitly assumed the electron density to be a function of the radial coordinate only (i.e., we assume the distribution of plasma surrounding the black hole to be spherically symmetric). Now, recalling that the plasma refractive index is given by \cite{Tsupko2013, Tsupko2021}
\begin{equation} \label{n-refr-index}
    n(r) = \sqrt{1 - \frac{\omega_{e}^{2}(r)}{\omega^{2}(r)}} = \sqrt{1 - A(r)\frac{\omega_e^2 (r)}{\omega_\infty^2}},
\end{equation}
we can rewrite the deflection angle as
\begin{equation}\label{deflection angle final}
    \hat{\alpha}(r_0) = 2\int_{r_0}^\infty \frac{\sqrt{B(r)}}{\sqrt{C(r)}\sqrt{\frac{C(r)}{C(r_0)}\frac{A(r_0)}{A(r)}\frac{n^2 (r)}{n^2 (r_0)}-1}}dr - \pi.
\end{equation}
An illustration of the physical scenario we have in mind is shown in Fig.\ 1.
\begin{figure}[t]
\includegraphics[width=8cm]{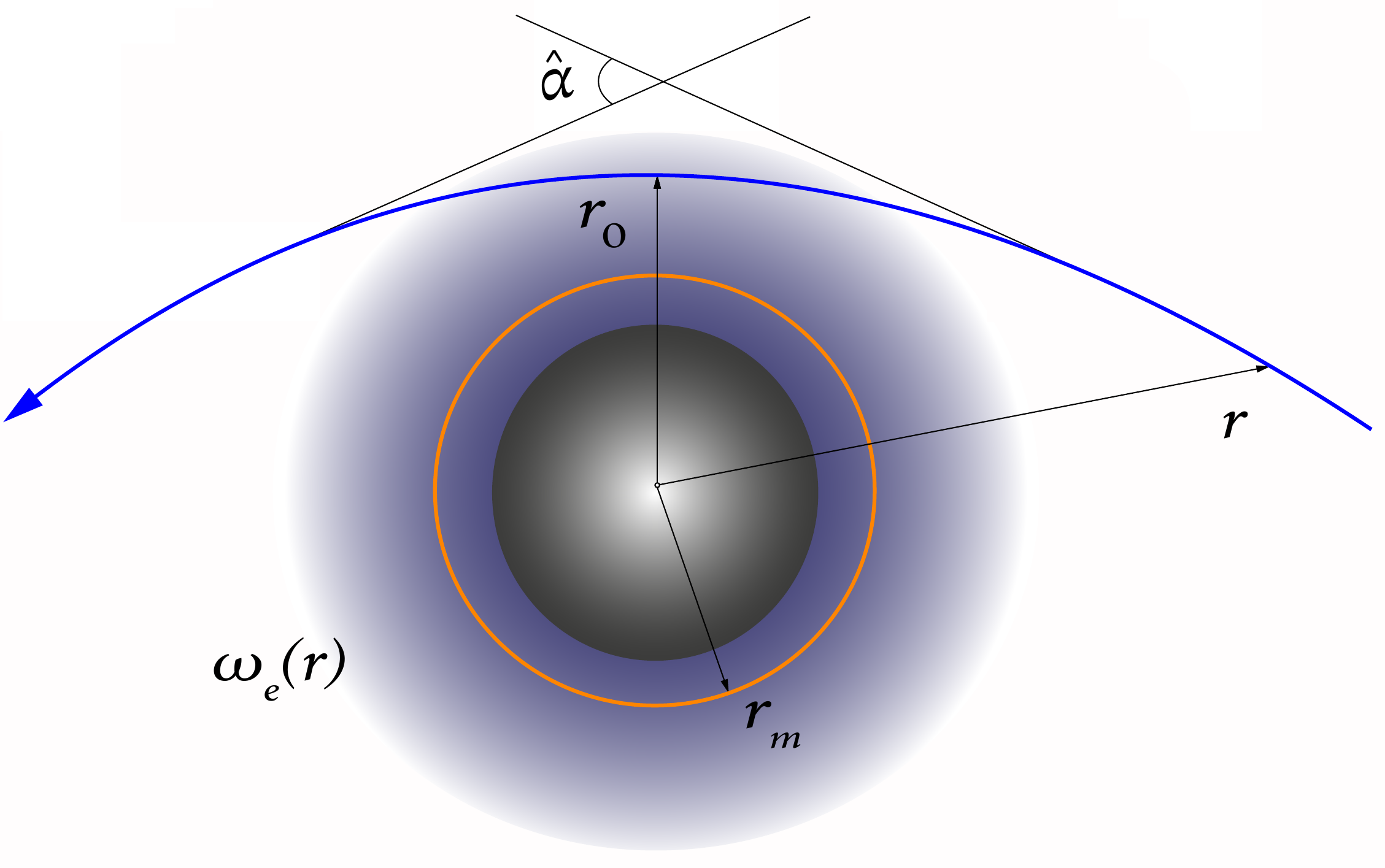}
\centering
\caption{Deflection angle $\hat{\alpha}$ of a photon moving nearby a very compact object surrounded by cold non-magnetized plasma whose frequency is denoted by $\omega_e(r)$. The value $r_0$ denotes the minimum value of the radial coordinate $r$ for this trajectory, while $r_m$ denotes the radius of the photon sphere. Inspired by Fig.\ 1 in Ref.\ \cite{Tsupko2021}.}
\end{figure}

In terms of the function $h(r)$, the photon sphere equation takes the simple form \cite{Perlick2015, Tsupko2021}
\begin{equation}\label{photon sphere equation}
    \frac{d}{dr}h^2 (r) = 0,
\end{equation}
where we recall that the photon sphere is the sphere covered by all unstable circular orbits of photons with any possible inclinations. Note that if the central object is not a black hole but another compact object, it is implied that its size is smaller than the size of the photon sphere.

\section{Strong deflection limit in the presence of plasma} \label{sec:sdlpl}

In this section, our goal is to perform a strong deflection limit analysis on Eq.\ \eqref{deflection angle final} adopting the formalism presented in Ref.\ \cite{Bozza2002}. This analysis aims to derive a general expression for the deflection angle in a plasma-filled background. The method developed in Ref.\ \cite{Bozza2002} applies to light rays that come close to the photon sphere, whose radius will be denoted by $r_{m}$. As $r_{0}$ approaches $r_{m}$, the deflection angle approaches infinity:\ photons can make one or several revolutions before flying off to infinity.

Before proceeding, let us notice that Eq.\ \eqref{deflection angle final} in Sec.\ \ref{sec:deflangwithpl} can be rewritten as
\begin{equation}\label{dangle}
    \hat{\alpha}(r_0) = 2\int_{r_0}^{\infty}\frac{\sqrt{B}}{\sqrt{C}\sqrt{\frac{C}{C_0}\frac{\mathcal{A}_0}{\mathcal{A}} - 1}}dr - \pi.
\end{equation}
where we defined $\mathcal{A} \coloneqq An^{-2}$ and, to avoid clutter of notation, we introduced the subscript $0$ which indicates that the function is evaluated at $r_0$. Moreover, we also omitted radial dependence. This simple redefinition of the coefficient $A$ allows us to immediately verify that in the two simplest limiting cases, $n = 1$ and $n = \text{constant}$, the results in vacuum \cite{Bozza2002} are recovered, with Eq.\ \eqref{dangle} above formally identical to Eq.\ (6) in Ref.\ \cite{Bozza2002}; we could then in principle write down all the equations of Sec.\ II in Ref.\ \cite{Bozza2002} with $\mathcal{A}$ in place of $A$. From the mathematical point of view, however, we find it convenient to introduce the procedure directly specifying the expressions in terms of the refractive index; by doing so, the calculation of the sole integral of the strong deflection limit procedure is indeed straightforward for most of the cases considered.

We start by requiring that the photon sphere equation admits at least one positive solution. The largest root of Eq.\ \eqref{photon sphere equation} gives the radius of the outermost photon sphere, $r_{m}$. Now, we proceed by substituting the integration variable $r$ in Eq.\ \eqref{deflection angle final} with a new variable, $z$, as
\begin{equation}
    z = \frac{A(r) - A_0}{1 - A_0},
\end{equation}
allowing us to write down the deflection angle as
\begin{equation}\label{alpha}
    \hat{\alpha}(r_0) = I(r_0) - \pi,
\end{equation}
where $I(r_0)$ is defined by the integral
\begin{equation}\label{initialintegral}
    I(r_0) \coloneqq \int_0^1 R(z,r_0)f(z,r_0)dz,
\end{equation}
with the two functions $R(z,r_0)$ and $f(z,r_0)$ given by
\begin{align}
    R(z,r_0) &\coloneqq \frac{2n_0 \sqrt{ABC_0}(1 - A_0)}{CA^\prime}, \label{R}
    \\
    f(z,r_0) &\coloneqq \frac{1}{\sqrt{A_0 n^2 - [(1 - A_0)z + A_0]\frac{C_0}{C}n_0^2}}. \label{f}
\end{align}
Notice that all functions without the subscript $0$ are evaluated at $r = A^{-1}\left[A_{0} + \left(1 - A_{0}\right)z\right]$. The function $R(z,r_0)$ is regular for all $z$ and $r_0$, while $f(z,r_0)$ diverges for $z \to 0$. To find out the order of the divergence, we expand the argument of the square root in $f(z,r_0)$ to the second order in $z$, obtaining
\begin{equation}\label{f_0}
    f_0(z,r_0) \coloneqq \frac{1}{\sqrt{\alpha(r_0)z + \beta(r_0)z^2}},
\end{equation}
where the coefficient $\alpha = \alpha(r_0)$ is defined as
\begin{equation}
    \alpha \coloneqq \frac{n_0^2 \left(1-A_0\right)}{C_0 A_0^\prime}\left[C_0^\prime A_0 + C_0 \left(2A_0 \frac{n_0^\prime}{n_0} -A_0^\prime\right)\right],
\end{equation}
while $\beta = \beta(r_0)$ as
\begin{multline}\label{betazero}
    \beta \coloneqq \frac{n_0^2 (1 - A_0)^2}{2C_0^2 A_0^{\prime 3}}\left[2C_0 C_0^\prime A_0^{\prime 2} \right.
    \\
    \left.
    \hspace{-1cm} + \hspace{1mm} A_0^\prime A_0 \left(C_0 C_0^{\prime \prime} - 2C_0^{\prime 2}\right) \right.
    \\
    \left.
    \hspace{0.6cm} - \hspace{1mm} C_0^2 A_0^{\prime \prime} A_0 \left(\frac{C_0^\prime}{C_0}  + 2\frac{n_0^\prime}{n_0}\right)\right]
    \\
    + \frac{A_0(1 - A_0)^2}{A_0^{\prime 2}}\left(n_0^{\prime 2} + n_0 n_0^{\prime \prime}\right).
\end{multline}
As we can see, when $\alpha(r_0)$ is non-zero, the leading order of the divergence in $f_0 (z,r_0)$ is $z^{-1/2}$, which can be integrated to give a finite result; on the other hand, when $\alpha(r_0)$ vanishes, the leading order of the divergence in $f_0 (z,r_0)$ is $z^{-1}$, which results in a divergent integral. Since we are interested in those trajectories whose inversion point is very close to the radius of the photon sphere $r_m$, we define a parameter $\delta \ll 1$ through the equation
\begin{equation}
    r_0 \coloneqq r_m (1 + \delta).
\end{equation}
We also notice that $\alpha(r_{0})$ vanishes at $\delta = 0$. Following Ref.\ \cite{Bozza2002}, we decompose the integral in Eq.\ \eqref{initialintegral} as
\begin{align}
    I(r_0) &= \int_0^1 R(0,r_m)f_0 (z,r_0)dz + \int_0^1 g(z,r_0)dz \nonumber \\
    &\coloneqq I_D (r_0) + I_R (r_0),
\end{align}
where the function $g(z,r_0)$ has been defined as
\begin{equation}\label{function g}
    g(z,r_0) \coloneqq R(z,r_0)f(z,r_0) - R(0,r_m)f_0 (z,r_0).
\end{equation}
As $\delta \to 0$, the integral $I_D (r_0)$ diverges, while $I_R (r_0)$ is regular (it is indeed given by the original integral with the divergence subtracted). The integral $I_{D}(r_{0})$ can be explicitly calculated, resulting in
\begin{equation}\label{I_D}
    I_D (r_0) = \frac{2R(0, r_m)}{\sqrt{\beta}}\log \left(\frac{\sqrt{\beta} + \sqrt{\alpha + \beta}}{\sqrt{\alpha}}\right).
\end{equation}
We proceed by expanding $\alpha$ up to $\mathcal{O}\left(\delta\right)$, obtaining
\begin{equation}\label{gammazero}
    \alpha = \frac{2\beta_m A_m^\prime r_m}{1-A_m}\delta + \mathcal{O}(\delta^2),
\end{equation}
where $\beta_m$ reads
\begin{multline}\label{betam}
    \beta_m = \frac{n_m\left(1-A_m\right)^2}{2C_m A_m^{\prime 2}}\left[n_m \left(C_m^{\prime \prime}A_m - C_m A_m^{\prime \prime}\right)\right.
    \\
    \left. + \left(3C_m^\prime A_m + C_m A_m^\prime \right)n_m^\prime + 2A_m C_m n_m^{\prime \prime}\right].
\end{multline}
To obtain \eqref{gammazero}, the photon sphere equation has been used. Also, we have introduced the subscript $m$ which indicates that the function is evaluated at $r_m$. Starting from these considerations, Eq.\ \eqref{I_D} is approximated as
\begin{equation}
    I_D (r_0) \simeq -a\log\delta(r_0) + b_D + \mathcal{O}(\delta),
\end{equation}
with $a$ and $b_D$ given by
\begin{align}
    a &\coloneqq \frac{R(0,r_m)}{\sqrt{\beta_m}}, \label{a}\\
    b_D &\coloneqq a\log\frac{2\left(1 - A_m\right)}{A_m^\prime r_m}, \label{b_D}
\end{align}
respectively. As for $I_{R}(r_{0})$, expanding it in powers of $\delta$ and considering only the first term of the expansion results in
\begin{equation}\label{regular term general}
    I_{R}(r_{0}) = \int_{0}^{1}g(z,r_{m})dz + \mathcal{O}\left(\delta\right) \coloneqq b_{R}.
\end{equation}
Putting it all together, we can finally write Eq.\ \eqref{alpha} as
\begin{equation}\label{deflection angle}
    \hat{\alpha}(r_0) = -a\log\delta(r_0) + b,
\end{equation}
where $a$ is defined by Eq.\ \eqref{a} while $b$ is given by
\begin{equation}\label{b}
    b \coloneqq b_D + b_R - \pi.
\end{equation}
This concludes our discussion on the strong deflection limit analysis in the presence of cold non-magnetized plasma. To summarize, the procedure consists of:
\begin{enumerate}[(i)]
    \item Solving Eq.\ \eqref{photon sphere equation} (or, alternatively, $\alpha(r_{0}) = 0$) to find the radius of the photon sphere.
    \item Computing $\beta_{m}$ from \eqref{betam} (or from \eqref{betazero} evaluated at $r_0 = r_m$) and the function $R(0,r_{m})$ from \eqref{R}.
    \item Computing the coefficient $b_{R}$ given by \eqref{regular term general} analytically or numerically, depending on the specific case.
    \item Computing the coefficients $a$ and $b$ from Eqs.\ \eqref{a} and \eqref{b}, respectively.
\end{enumerate}
As we will see in the next section, the critical steps are the calculation of the photon sphere radius and of the coefficient $b_{R}$.

The deflection angle can also be written in terms of the impact parameter, here denoted by $u$, which in the presence of plasma is defined by the equation \cite{Perlick-Tsupko-2024}
\begin{equation}\label{impactpar}
    u = \frac{n_0}{n_\infty}\sqrt{\frac{C_0}{A_0}}.
\end{equation}
with $n_\infty \coloneqq n(r \to \infty)$. As $\delta \to 0$ (i.e., as $r_0 \to r_m$), also the impact parameter must be close to its minimum $u_m$; we thus define the parameter $\varepsilon \ll 1$ through the equation
\begin{equation}
    u \coloneqq u_m (1 + \varepsilon), \quad u_{m} = \frac{n_m}{n_\infty}\sqrt{\frac{C_m}{A_m}}.
\end{equation}
Now, expanding Eq.\ \eqref{impactpar} around $\delta = 0$, we find
\begin{equation}
    u - u_m = \Tilde{c} \, r_m^2 \delta^2 = u_m \varepsilon,
\end{equation}
where $\Tilde{c}$ is defined to be
\begin{equation}
    \Tilde{c} \coloneqq \frac{\beta_m A_m^{\prime 2}\sqrt{C_m}}{2n_\infty n_m A_m^{3/2}(1 - A_m)^2}.
\end{equation}
In terms of $\epsilon$, the deflection angle can be written as
\begin{equation}\label{deflection angle imppar}
    \hat{\alpha}(u) = -\Bar{a}\log\varepsilon(u) + \Bar{b},
\end{equation}
where the coefficients $\Bar{a}$ and $\Bar{b}$ are given by
\begin{align}
    \Bar{a} &\coloneqq \frac{a}{2} = \frac{R(0,r_m)}{2\sqrt{\beta_m}}, \\
    \Bar{b} &\coloneqq -\pi + b_R + \Bar{a}\log\frac{2\beta_m}{n_m^2 A_m},
\end{align}
respectively.

\section{Schwarzschild lensing in the presence of plasma} \label{sec:schwlensing}
In the previous section, building upon Ref.\ \cite{Bozza2002}, the strong deflection limit analysis in a static, asymptotically flat and spherically symmetric spacetime has been extended to include the presence of plasma. In particular, an analytic expression for the deflection angle has been derived, both in terms of the closest approach distance $r_0$, Eq.\ \eqref{deflection angle}, and in terms of the impact parameter $u$, Eq.\ \eqref{deflection angle imppar}. As anticipated, in this section we will specialize our discussion to the case of a Schwarzschild black hole, the simplest spherically symmetric vacuum solution of the Einstein field equations. 

For convenience, we define the Schwarzschild radius as the unit of measure of distances; then, in Schwarzschild coordinates, the metric coefficients take the form
\begin{align}
    A(r) &= 1 - \frac{1}{r}, \\
    B(r) &= \left(1 - \frac{1}{r}\right)^{-1}, \\
    C(r) &= r^2.
\end{align}
For the reader's convenience, we also recall the expression for the plasma refractive index, that is
\begin{equation}\label{refractive index plasma}
    n(r) = \sqrt{1 - \frac{\omega_{e}^{2}(r)}{\omega^{2}(r)}} = \sqrt{1 - A(r)\frac{\omega_e^2 (r)}{\omega_\infty^2}}.
\end{equation}
We remind the reader that $\omega(r)$ denotes the photon frequency measured by a static observer, $\omega_\infty$ is the photon frequency at infinity, and $\omega_e(r)$ represents the plasma frequency to be specified.

In what follows, after reproducing the already known result for a homogeneous density profile \cite{Tsupko2013} in Sec.\ \ref{subsec:hompl}, we will consider non-uniform plasma, deriving exact or approximate expressions depending on the specific density profile chosen. We will conclude Sec.\ \ref{sec:schwlensing} by presenting results for a plasma with arbitrary radial power-law density profile.

To avoid clutter of notation, we will specify the dependencies of the various quantities only when strictly necessary, e.g.\ when writing down the final expression for the deflection angle.

\subsection{Homogeneous plasma} \label{subsec:hompl}
As anticipated, before discussing the more realistic scenario where the Schwarzschild black hole is surrounded by non-uniform plasma, here, we consider the case of a homogeneous plasma: $\omega_{e}(r) = \omega_{e} = \text{constant}$. Even if this case has already been analyzed in the literature \cite{Tsupko2013}, it is worth it to show how the same expression for the deflection angle can be obtained within the formalism presented in this paper.

Defining $\Tilde{\omega}^2 \coloneqq \omega_e^2 / \omega_\infty^2$, the two functions $R(z,r_{0})$ and $f(z,r_{0})$ introduced earlier in Eqs.\ \eqref{R} and \eqref{f} read
\begin{align}
    R(z,r_{0}) &= R(r_{0}) = 2n_{0} = 2\sqrt{1 - \left(1 -\frac{1}{r_{0}}\right)\Tilde{\omega}^2}, \\
    f(z,r_{0}) &= \frac{1}{\sqrt{\alpha z + \beta z^2 - \gamma z^3}},
\end{align}
where the coefficients $\alpha, \beta$ and $\gamma$ are given by
\begin{align}
    \alpha = \alpha(r_{0}) &= 2 - \frac{3}{r_0} - 2\Tilde{\omega}^2 \left(1 - \frac{2}{r_0} + \frac{1}{r_0^2}\right), \\
    \beta = \beta(r_{0}) &= \frac{3}{r_0} - 1 + \Tilde{\omega}^2 \left(1 - \frac{4}{r_0} + \frac{3}{r_0^2}\right), \\
    \gamma = \gamma(r_{0}) &= \frac{1}{r_0}\left[1 - \Tilde{\omega}^2 \left(1 - \frac{1}{r_0}\right)\right].
\end{align}
If $\omega_{e} = 0$, it is immediate to verify that the above expressions reduce to Eqs. (43) and (44) of Ref. \cite{Bozza2002}. Now, the radius of the photon sphere can be found by setting $\alpha(r_{0}) = 0$, resulting in
\begin{equation}
    r_0^2 - \left(\frac{3 - 4\Tilde{\omega}^2}{2 - 2\Tilde{\omega}^2}\right)r_0 - \frac{\Tilde{\omega}^2}{1 - \Tilde{\omega}^2} = 0.
\end{equation}
The solution of the above equation which, as $\omega_{e} \to 0$, reduces to the well-known vacuum result (i.e., 3/2) is
\begin{equation}\label{radius photon sphere plasma}
    r_{m} = \frac{\sqrt{9 - 8\Tilde{\omega}^2} - 4\Tilde{\omega}^2 + 3}{4\left(1 - \Tilde{\omega}^2\right)}.
\end{equation}
In order to make contact with Ref. \cite{Tsupko2013}, we define 
\begin{equation}
    x \coloneqq \sqrt{1 - \frac{8}{9}\Tilde{\omega}^2} = \sqrt{1 - \frac{8\omega_e^2}{9\omega_\infty^2}}.
\end{equation}
In terms of $x$, Eq. \eqref{radius photon sphere plasma} can be rewritten as
\begin{equation}\label{radius photon sphere final}
    r_{m} = 3\frac{1 + x}{1 + 3x}.
\end{equation}
Consequently, from Eq.\ \eqref{a}, we compute the coefficient $a$, finding
\begin{equation}
    a = 2\sqrt{\frac{1 - \Tilde{\omega}^2 + \frac{\Tilde{\omega}^2}{r_m}}{\frac{3}{r_m} - 1 + \Tilde{\omega}^2 - \frac{4\Tilde{\omega}^2}{r_m} + \frac{3\Tilde{\omega}^2}{r_m^2}}} = 2\sqrt{\frac{1 + x}{2x}}.
\end{equation}
The coefficient $\beta$ has been read off from the expansion of the denominator of $f(z,r_{0})$, and then it has been evaluated at $r_0 = r_m$. For consistency, one can immediately check that the same expression for $\beta_{m}$ can be obtained from Eq.\ \eqref{betam}. Let us now consider the other two coefficients to be calculated, $b_D$ and $b_R$. The former gives
\begin{equation}
    b_D = -2\sqrt{\frac{1 + x}{2x}}\log\frac{1}{2}.
\end{equation}
Concerning the latter, we have
\begin{align}\label{regular term}
    b_R &= 2n_m \int_0^1 \left(\frac{1}{\sqrt{\beta_m z^2 - \gamma_m z^3}} - \frac{1}{\sqrt{\beta_m}z}\right)dz \nonumber \\
    &= -a\left[\log\left(\frac{\sqrt{1 - \frac{\gamma_m}{\beta_m}} + 1}{1 - \sqrt{1 - \frac{\gamma_m}{\beta_m}}}\right) + \log\frac{\gamma_m}{4\beta_m}\right] \nonumber \\
    &= -a\log\left[\frac{\left(\sqrt{3x - 1} + \sqrt{6x}\right)^2}{24x}\right].
\end{align}
Putting it all together, we can finally write the deflection angle in terms of $x$ as
\begin{equation}\label{deflection angle hp}
    \hat{\alpha}(r_0, x) = -2\sqrt{\frac{1 + x}{2x}}\log\left(z_1 (x) \delta(r_{0},x)\right) - \pi,
\end{equation}
where the quantity $z_1$ is defined by
\begin{equation}
    z_1 (x) \coloneqq \frac{9x - 1 + 2\sqrt{6x(3x - 1)}}{48x},
\end{equation}
while $\delta(r_{0},x)$ is given by
\begin{equation}
    \delta(r_{0},x) = \frac{r_0}{r_m (x)} - 1.
\end{equation}
Eq.\ (\ref{deflection angle hp}) agrees with Eq. (79) in Ref.\ \cite{Tsupko2013}. 

In terms of the impact parameter $u$, the deflection angle results in
\begin{equation}\label{daimp}
    \hat{\alpha}(u, x) = -\Bar{a}(x)\log\varepsilon(u,x) + \Bar{b}(x),
\end{equation}
where $\varepsilon(u,x)$, $\Bar{a}(x)$ and $\Bar{b}(x)$ read
\begin{align}
    \varepsilon(u,x) &= \frac{u}{u_m(x)} - 1,\\
    \Bar{a}(x) &= \sqrt{\frac{1 + x}{2x}}, \\
    \Bar{b}(x) &= -\Bar{a}(x)\log\left(\frac{2z_1^2 (x)}{3x}\right) - \pi,
\end{align}
respectively, with $u_m(x)$ given by
\begin{equation}
   u_m(x) = \sqrt{\frac{3(1 + x)}{3x - 1}}r_m. 
\end{equation}
Eq.\ \eqref{daimp} is in agreement with Eq.\ (88) in Ref.\ \cite{Tsupko2013}.

As we have seen, the homogeneous plasma case can be solved exactly: the radius of the photon sphere can be indeed found without relying on any approximation and, moreover, no numerical integration is needed to calculate the coefficient $b_R$.

In Fig.\ 2, the comparison between the exact deflection angle (found by numerical integration) and the one in the strong deflection limit is plotted, showing excellent agreement for $r_{0}$ close to the photon sphere $r_m$.
\begin{figure}[t]
\includegraphics[width=8cm]{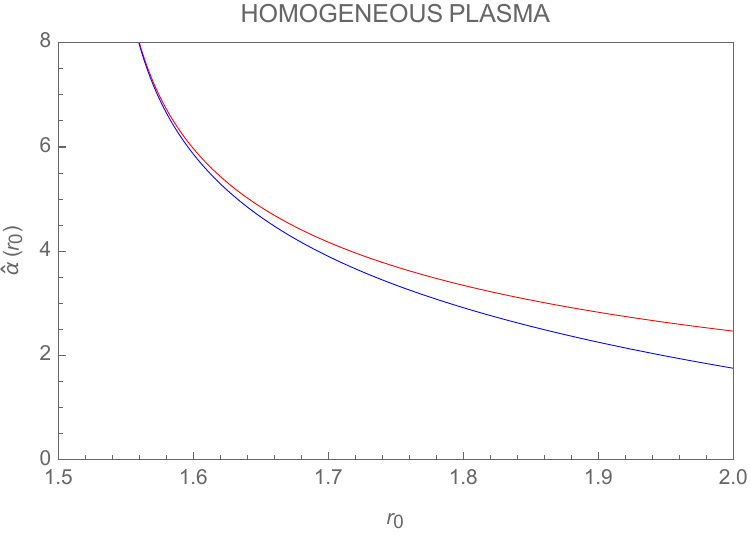}
\centering
\caption{Deflection angle in Schwarzschild spacetime surrounded by homogeneous plasma as a function of the minimum radial coordinate $r_0$, Sec.\ \ref{subsec:hompl}. The red curve is the result of a numerical calculation, while the blue curve represents the deflection angle calculated using the strong deflection limit formula \eqref{deflection angle hp}. We set the value $\Tilde{\omega}^2 = 0.2$. As expected, the agreement is excellent when $r_0$ is near the photon sphere radius.}
\end{figure}

\subsection{Plasma with density profile $N(r) \propto r^{-1}$}
\label{subsec:nonuniformpl1}
Let us now consider a non-uniform plasma with comoving number density of the form
\begin{equation}
    N(r) = \frac{N_{c_1}}{r},
\end{equation}
where $N_{c_1}$ is a constant. Consequently, the plasma frequency can be rewritten as
\begin{equation}
    \omega_{e} (r) = \sqrt{\frac{4\pi e^2 N_{c_1}}{m_{e}r}}.
\end{equation}
For further simplicity, we introduce the constant
\begin{equation}\label{constantk}
	k \coloneqq \frac{4\pi e^2 N_{c_1}}{\omega_\infty^2 m_{e}},
\end{equation}
which fully characterizes the magnitude of plasma influence compared to the vacuum case. A larger value of this coefficient indicates a greater influence of plasma. If $N_{c_1} = 0$ or $\omega_\infty \to \infty$, we have $k = 0$, and the vacuum case is recovered.

The functions $R(z,r_{0})$ and $f(z,r_{0})$ read
\begin{align}
    R(z,r_{0}) &= R(r_{0}) = 2n_{0} = 2\sqrt{1 - \left(1 -\frac{1}{r_{0}}\right)\frac{k}{r_{0}}}, \\
    f(z,r_{0}) &= \frac{1}{\sqrt{\alpha z + \beta z^2 - \gamma z^3}}, \label{f1}
\end{align}
where the coefficients $\alpha, \beta$ and $\gamma$ are given by
\begin{align}
    \alpha &= 2 - \frac{3 + k}{r_0} + \frac{k}{r_0^2}\left(2 - \frac{1}{r_0}\right), \\
    \beta &= \frac{3 + k}{r_0} - 1 - \frac{k}{r_0^2}\left(3 -\frac{2}{r_0}\right), \\
    \gamma &= \frac{1}{r_0} - \frac{k}{r_0^2}\left(1 - \frac{1}{r_0}\right).
\end{align}
As before, the radius of the photon sphere can be found from the equation $\alpha = 0$, leading to
\begin{equation}\label{rm1}
    r_{m} = \frac{1}{6}\left[k + 3 + g(k) + \frac{\left(k - 3\right)^2}{g(k)}\right],
\end{equation}
where the function $g(k)$ is defined as
\begin{equation}
    g(k) \coloneqq \sqrt[3]{\left(k - 3\right)^3 + 54 + 6\sqrt{3}\sqrt{\left(k - 3\right)^3 + 27}}.
\end{equation}
Setting $k = 0$ gives $g(0) = 3$ and $r_{m} = 3/2$, as expected. The coefficient $a$ results in
\begin{equation}\label{coefficienta1}
    a = 2\sqrt{\frac{1 - \left(1 - \frac{1}{r_m}\right)\frac{k}{r_m}}{\frac{3 + k}{r_m} - 1 -k\left(\frac{3}{r_m^2} -\frac{2}{r_m^3}\right)}},
\end{equation}
with $r_{m}$ given by Eq. \eqref{rm1}. As for the coefficients $b_D$ and $b_R$, no complications arise with respect to the homogeneous case, finding
\begin{align}
    b_{D} &= a\log2, \label{bd1} \\
    b_{R} &= -a \log\left[\frac{1 + \sqrt{1 - \frac{1}{3 - r_m}}}{4\left(3 - r_m\right)\left(1 - \sqrt{1 - \frac{1}{3 - r_m}}\right)}\right], \label{br1}
\end{align}
respectively. We can thus write down the formal expression for the deflection angle as
\begin{equation}\label{daq1}
    \hat{\alpha}(r_0, k) = -a(k)\log\delta(r_0 ,k) + b(k),
\end{equation}
where $\delta(r_0, k)$ is
\begin{equation}
    \delta(r_0, k) = \frac{r_0}{r_m (k)} - 1.
\end{equation}
The quantities $r_m (k)$ and $a(k)$ are given by Eqs.\ \eqref{rm1} and \eqref{coefficienta1}, respectively, and $b(k) = b_D (k) + b_R (k) - \pi$, where $b_D (k)$ and $b_R (k)$ are given by Eqs.\ \eqref{bd1} and \eqref{br1}, respectively. In Fig.\ 3, the comparison between the numerical calculation of the deflection angle and the one computed in the strong deflection limit is shown.
\begin{figure}[t]
\includegraphics[width=8cm]{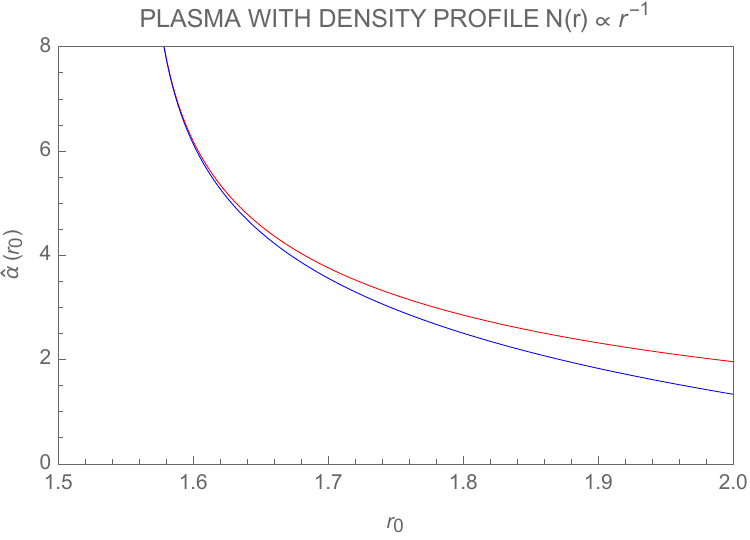}
\centering
\caption{Deflection angle in Schwarzschild spacetime surrounded by non-uniform plasma with density profile of the form $N(r) \propto r^{-1}$; see Sec.\ \ref{subsec:nonuniformpl1}. The red curve is the result of a numerical calculation while the blue curve represents the deflection angle in the strong deflection limit, see Eq.\ \eqref{daq1}. In both cases, the constant $k$ introduced in (\ref{constantk}) is chosen to be equal to $k = 1$.}
\end{figure}
\begin{figure}[t]
\includegraphics[width=8cm]{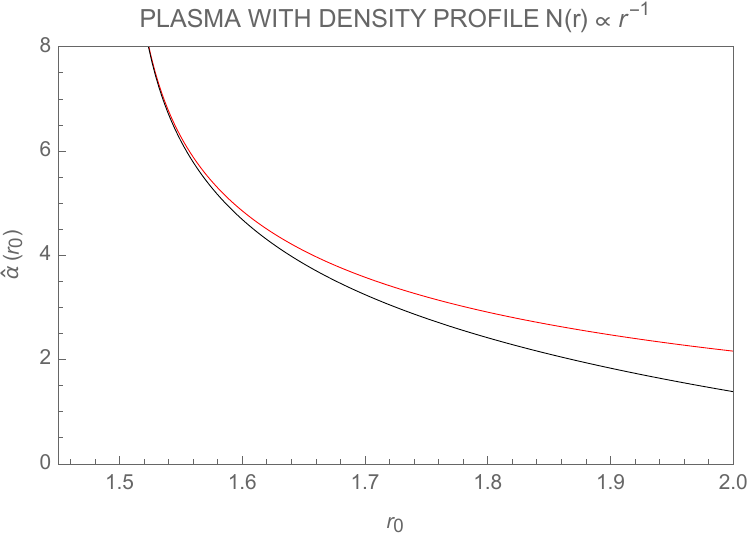}
\centering
\caption{Deflection angle in Schwarzschild spacetime surrounded by non-uniform plasma with density profile of the form $N(r) = N_{0}r^{-1}$. The red line is the result of a numerical calculation while the black line represents the deflection angle both in the strong deflection limit and in the low-density plasma approximation, Eq.\ \eqref{daq1approx}. In both cases we set $k = 0.1$.}
\end{figure}
As we can notice, also the case of a non-uniform plasma with a density profile of the form $N(r) \propto 1/r$ can be solved exactly. However, for later convenience, we can specialize our formulas to the realistic case when the plasma frequency is much smaller than the photon frequency; to implement this approximation, we rewrite the refractive index as
\begin{equation}
    n(r) = \sqrt{1 - \epsilon\frac{\omega_e^2 (r)}{\omega^2(r)}} = \sqrt{1 - \epsilon\left(1 - \frac{1}{r}\right)\frac{k}{r}}.
\end{equation}
Above, we have introduced a book-keeping parameter, denoted by $\epsilon$, which will be set to unity after linearizing all equations with respect to $\epsilon$.
The function $f(z,r_{0})$ assumes of course the same form as in \eqref{f1}, but now the coefficients $\alpha, \beta$ and $\gamma$ are given by
\begin{align}
    \alpha &= 2 - \frac{3 + \epsilon k}{r_0} + \frac{\epsilon k}{r_0^2}\left(2 - \frac{1}{r_0}\right), \\
    \beta &= \frac{3 + \epsilon k}{r_0} - 1 - \frac{\epsilon k}{r_0^2}\left(3 -\frac{2}{r_0}\right), \\
    \gamma &= \frac{1}{r_0} - \frac{\epsilon k}{r_0^2}\left(1 - \frac{1}{r_0}\right).
\end{align}
By solving the equation $\alpha = 0$ and then linearizing with respect to $\epsilon$, we obtain
\begin{equation}\label{radius photon sphere first order}
    r_m \simeq \frac{3}{2} + \epsilon \frac{k}{18}.
\end{equation}
In this approximation, the coefficient $a$ becomes
\begin{equation}\label{coefficient a h=1}
    a \simeq 2\left(1 - \frac{k}{27}\right),
\end{equation}
where in the last step we again linearized with respect to $\epsilon$ and then set $\epsilon = 1$. Applying the same strategy to the regular term, we find
\begin{equation}
    b_{R} \simeq b_{R,0} + b_{R,1}k,
\end{equation}
where $b_{R,0}$ and $b_{R,1}$ are written as
\begin{align}
    b_{R,0} &= 2\log\left[6\left(2 - \sqrt{3}\right)\right] \approx 0.95, \label{vacuum b_R} \\
    b_{R,1} &= -\frac{2}{27}\left(3\sqrt{3} - 3 + \frac{b_{R,0}}{2} \right) \approx -0.20, \label{br11} 
\end{align}
respectively. Finally, the coefficient $b_{D}$ is given by
\begin{equation}
    b_{D} \simeq 2\left(1 - \frac{k}{27}\right)\log2.
\end{equation}
In the low-density plasma approximation, the deflection angle can then be explicitly written as
\begin{multline}\label{daq1approx}
    \hat{\alpha}(r_{0},k) \simeq -\hspace{0.5mm}2\left(1 - \frac{k}{27}\right)\log\delta(r_0, k) 
    \\
    \hspace{0.2cm} + 2\left(1 - \frac{k}{27}\right)\log\left[12\left(2 - \sqrt{3}\right)\right]
    \\
    -\frac{2k}{9}\left(\sqrt{3} - 1\right) - \pi.
\end{multline}
As expected, by setting $k = 0$ in the above formula, the well-known result in the absence of plasma \cite{Bozza2002} is recovered. In Fig.\ 4, Eq.\ \eqref{daq1approx} is plotted and compared against the numerical calculation. It is worth mentioning that in Fig.\ 4, the constant $k$ has been set to $0.1$, not unity as in Figs.\ 2 and 3. 
Indeed, in Fig.\ 3, we are  plotting the numerical calculation of $\alpha(r_0)$ against both the strong deflection limit and the low-density plasma limit. It is therefore important to verify the approximation's accuracy and determine its validity range.

Computing $u_m, \bar{a}$ and $\Bar{b}$, we can also express the deflection angle in terms of the impact parameter as
\begin{multline}\label{daimp1}
    \hat{\alpha}(u,k) \simeq -\left(1 - \frac{k}{27}\right)\log\varepsilon(u, k) \\
    \hspace{1.05cm}+ \log\left[216 \left(7 - 4\sqrt{3}\right)\right] - \frac{2k}{9}\left\{\sqrt{3} - 1 \right. \\
    \left. +\hspace{0.5mm}\frac{1}{6}\log6 + \frac{1}{3}\log\left[6\left(2 - \sqrt{3}\right)\right]\right\} - \pi,
\end{multline}
with $\varepsilon(u,k)$ given by
\begin{equation}
    \varepsilon(u,k) = \frac{u}{u_m (u,k)} - 1 = \frac{2\sqrt{3}u}{9 - k} - 1. 
\end{equation}

\subsection{Plasma with density profile $N(r) \propto r^{-2}$}
\label{subsec:nonuniformpl2}
The plasma number density is now given by
\begin{equation}
    N(r) = \frac{N_{c_2}}{r^{2}}.
\end{equation}
This case is particularly interesting because with this choice of the plasma distribution, the radius $r_m$ of the photon sphere in the presence of plasma is exactly equal to the radius of the photon sphere in vacuum, which in our units is $3/2$, see Ref.\ \cite{Perlick2015}.

The plasma frequency and the refractive index are now given by the expressions
\begin{align}
    \omega_{e} (r) &= \sqrt{\frac{4\pi e^2 N_{c_2}}{m_{e}r^2}}, \\
    n(r) &= \sqrt{1 - \left(1 - \frac{1}{r}\right)\frac{k}{r^2}}, 
\end{align}
respectively. It is important to note that the constant $k$ introduced above is defined as
\begin{equation}
k \coloneqq \frac{4\pi e^2 N_{c_2}}{\omega_\infty^2 m_{e}}.
\end{equation}
Although it is different from the one introduced in the previous section ($N_{c1} \ne N_{c2}$), we do not change its label to avoid cluttering the notation.

The case considered in this section turns out to be surprisingly simple; this is due to the fact that the function $f(z,r_{0})$ has the same form as the one in the absence of plasma \cite{Bozza2002}, namely 
\begin{equation}
    f(z,r_{0}) = \frac{1}{\sqrt{\left(2 - \frac{3}{r_0}\right)z + \left(\frac{3}{r_0} - 1\right)z^2 - \frac{z^3}{r_0}}},
\end{equation}
from which we read off the coefficients $\alpha$ and $\beta$:
\begin{align}
    \alpha &= 2 - \frac{3}{r_{0}}, \\
    \beta &= \frac{3}{r_{0}} - 1.
\end{align}
By imposing $\alpha = 0$ we find the radius of the photon sphere, $r_{m} = 3/2$, which immediately leads to $\beta_{m} = 1$. The coefficients $a, b_{R}$ and $b_{D}$ turn out to be
\begin{align}
    a &= 2\sqrt{1 - \frac{4k}{27}}, \\
    b_{R} &= a\log\left[6\left(2 - \sqrt{3}\right)\right], \\
    b_{D} &= a\log2,
\end{align}
respectively. The deflection angle is thus given by
\begin{multline}\label{da2exact}
    \hat{\alpha}(r_{0},k) = -2\sqrt{1 - \frac{4k}{27}}\log\delta(r_0) 
    \\
    + 2\sqrt{1 - \frac{4k}{27}}\log\left[12\left(2 - \sqrt{3}\right)\right] - \pi.
\end{multline}

The low-density plasma version of Eq.\ \eqref{da2exact} reads
\begin{multline}
    \hat{\alpha}(r_{0},k) \simeq -2\left(1 - \frac{2k}{27}\right)\log\delta(r_0) 
    \\
    + 2\left(1 - \frac{2k}{27}\right)\log\left[12\left(2 - \sqrt{3}\right)\right] - \pi.
\end{multline}
For later convenience, we also write down the result we obtained for the regular term. Using the same notation as before, we can write $b_R \simeq b_{R,0} + b_{R,1}k$, with $b_{R,0}$ given by Eq.\ (\ref{vacuum b_R}) and
\begin{equation}
    b_{R,1} = -\frac{4}{27}\log\left[6\left(2 - \sqrt{3}\right)\right] \approx -0.07. \label{br12}
\end{equation}
In terms of $u$, the deflection angle instead results in
\begin{multline}\label{daimp2}
    \hat{\alpha}(u, k) \simeq -\left(1 - \frac{2k}{27}\right)\log\varepsilon(u,k)
    \\
    \hspace{0.2cm}+ \frac{4k}{27}\left\{1 - \log\sqrt{6} -\log\left[6\left(2 - \sqrt{3}\right)\right] \right\}
    \\
    + \log\left[216 \left(7 - 4\sqrt{3}\right)\right] - \pi,
\end{multline}
with $\varepsilon(u,k)$ given by
\begin{equation}
    \varepsilon(u,k) = \frac{u}{u_m (u,k)} - 1 = \frac{6\sqrt{3}u}{27 - 2k} - 1. 
\end{equation}

\subsection{Plasma with density profile $N(r) \propto r^{-3}$}
\label{subsec:nonuniformpl3}
The next scenario that one could hope to solve exactly is the one in which the Schwarzschild black hole is surrounded by a non-uniform plasma with number density $N(r)$ proportional to $1/r^{3}$. However, the first difficulty arises when we write down the photon sphere equation, which is not analytically solvable in this case. Therefore, we are forced to rely on the low-density plasma approximation from the very beginning. When the plasma frequency is significantly smaller than the photon frequency, we can linearize the equation for the photon sphere around the corresponding value for light rays in vacuum. For details, the reader may refer to Sec.\ V of Ref. \cite{Perlick2015}. In the case at hand, such linearization procedure results in
\begin{equation}\label{radius photon sphere h=3}
    r_{m} \simeq \frac{3}{2} - \epsilon \frac{2k}{81}.
\end{equation}
Before proceeding, we also write down the expression for the plasma frequency, that is
\begin{equation}
    \omega_{e} (r) = \sqrt{\frac{4\pi e^2 N_{c_3}}{m_{e}r^3}}.
\end{equation}
The function $f(z,r_{0})$ assumes a slightly more complicated form with respect to the previous cases, namely
\begin{equation}
    f(z,r_{0}) = \frac{1}{\sqrt{\alpha z + \beta z^2 - \gamma z^3 + \delta z^4}},  
\end{equation}
where the coefficients $\alpha, \beta, \gamma$ and $\delta$ read
\begin{align}
    \alpha &= 2 - \frac{3}{r_0} + \frac{\epsilon k}{r_0^3}\left(1 - \frac{2}{r_0} + \frac{1}{r_0^2}\right), \\
    \beta &= \frac{3}{r_0} - 1 - \frac{\epsilon k}{r_0^3}\left(2 - \frac{5}{r_0} + \frac{3}{r_0^2}\right), \label{beta h=3} \\
    \gamma &= \frac{1}{r_0} - \frac{\epsilon k}{r_0^3}\left(1 - \frac{4}{r_0} + \frac{3}{r_0^2}\right), \\
    \delta &= \frac{\epsilon k}{r_0^4}\left(1 - \frac{1}{r_0}\right). \label{eta h=3}
\end{align}
The constant $k$ above is defined as
\begin{equation}
    k \coloneqq \frac{4\pi e^2 N_{c_3}}{\omega_\infty^2 m_{e}}.
\end{equation}
As for the coefficient $a$, we have
\begin{align}
    a &= 2\sqrt{\frac{1 - \epsilon\left(1 - \frac{1}{r_m}\right)\frac{k}{r_m}}{\frac{3}{r_m} - 1 - \frac{\epsilon k}{r_m^3}\left(2 - \frac{5}{r_m} + \frac{3}{r_m^2}\right)}} \nonumber \\
    &\simeq 2\left(1 - \frac{16k}{243}\right),
\end{align}
where we linearized with respect to $\epsilon$ and finally set $\epsilon = 1$, as before. The next step is to compute the regular term, $b_{R}$; as it turns out, also in this case the integration can be easily performed, leading to
\begin{multline}\label{regular term h=3}
    b_{R} = a\left[2\arctanh{\left(\frac{\sqrt{\delta_m} - \sqrt{\beta_m - \gamma_m + \delta_m}}{\sqrt{\beta_m}}\right)}\right. \\
    \left. +\hspace{0.5mm} \log\left(\frac{4\beta_m}{\gamma_m + 2\sqrt{\beta_m \delta_m}}\right)\right],
\end{multline}
with $\beta_m > 0, \gamma_m < 1$ and $\beta_m > \gamma_m$. Now, by inserting Eqs.\ (\ref{beta h=3})--(\ref{eta h=3}) in Eq.\ (\ref{regular term h=3}), linearizing with respect to $\epsilon$ and finally setting $\epsilon = 1$, leads to
\begin{equation}
    b_R \simeq b_{R,0} + b_{R,1}k,
\end{equation}
with $b_{R,0}$ given by (\ref{vacuum b_R}) and
\begin{equation}
    b_{R,1} = -\frac{16}{243}\left[2\sqrt{3} - 5 + b_{R,0}\right] \approx 0.04. \label{br13}
\end{equation}
The coefficient $b_D$ is instead given by
\begin{equation}
    b_D \simeq 2\left(1 - \frac{16k}{243}\right)\log2,
\end{equation}
again after the linearization procedure. Putting it all together, we can write the deflection angle as
\begin{multline}\label{daq3}
    \hat{\alpha}(r_{0},k) \simeq -\hspace{0.5mm}2\left(1 - \frac{16k}{243}\right)\log\delta(r_0 ,k) 
    \\
    \hspace{1cm}+ 2\left(1 - \frac{16k}{243}\right)\log\left[12\left(2 - \sqrt{3}\right)\right]
    \\
    -\frac{16k}{243}\left(2\sqrt{3} - 5\right) - \pi.
\end{multline}

In summary, we have been able to obtain results in this case without relying on numerical procedures. However, due to the impossibility of solving the photon sphere equation exactly, we had to specialize all the expressions to the case where the plasma frequency is much smaller than the photon frequency.
As before, we also write down the deflection angle as a function of $u$, resulting in
\begin{multline}\label{daimp3}
    \hat{\alpha}(u,k) \simeq \left(1 - \frac{16k}{243}\right)\log\varepsilon(u,k)
    \\
    \hspace{-0.5cm} + \log\left[216 \left(7 - 4\sqrt{3}\right)\right] - \frac{16k}{243}\left\{2\sqrt{3} \right.
    \\
    \left. + \log6 - \frac{15}{2} + 2\log\left[6\left(2 - \sqrt{3}\right)\right] \right\} - \pi,
\end{multline}
with $\varepsilon(u,k)$ given by
\begin{equation}
    \varepsilon(u,k) = \frac{u}{u_m (u,k)} - 1 = \frac{18\sqrt{3}u}{81 - 4k} - 1. 
\end{equation}

\subsection{General case: $N(r) \propto r^{-q}$, $q > 0$}
\label{subsec:nonuniformplh}
The outlined procedure is by now clear. Here we generalize the results presented in the previous sections to the case of a non-homogeneous plasma with number density $N(r) \propto r^{-q}$, with $q > 0$.
In this case, the radius of the photon sphere and the refractive index read
\begin{align}\label{rmq}
    r_m &\simeq \frac{3}{2} + \epsilon \frac{2^{q - 1}}{3^{q + 1}}\left(1 - \frac{q}{2}\right)k, \\
    n(r) &= \sqrt{1 - \epsilon\left(1 - \frac{1}{r}\right)\frac{k}{r^{q}}},
\end{align}
respectively. In the general case we are considering here, the constant $k$ is defined as
\begin{equation}
    k \coloneqq \frac{4\pi e^2 N_{c_q}}{\omega_\infty^2 m_{e}}.
\end{equation}

Proceeding as before, i.e., reading off the coefficient $\beta$ from the function $f(z,r_0)$ is not possible anymore; this is simply due to the fact that the exponent in the power law, $q$, can be any real number:~the function $f(z,r_0)$ cannot be expanded in powers of $z$. We then calculate the coefficient $\beta_m$ directly evaluating \eqref{betazero} at $r_0 = r_m$. By doing so, and applying the linearization procedure introduced in the previous sections, the coefficients $a$ and $b_D$ turn out to be 
\begin{figure}[t!]
\includegraphics[width=8cm]{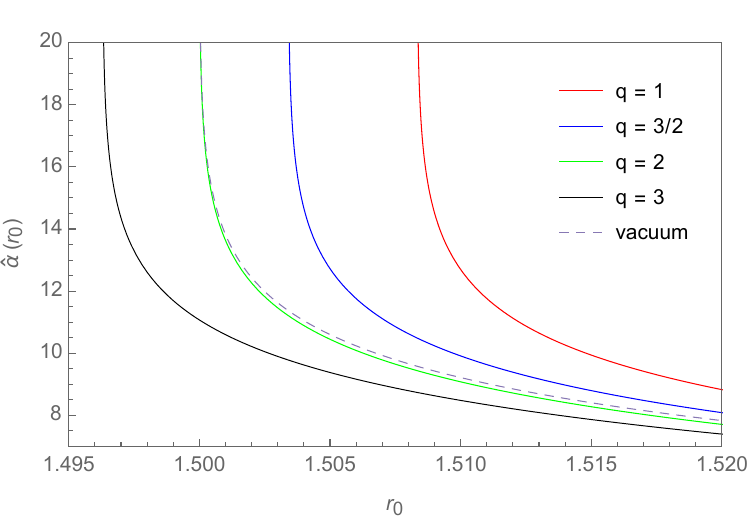}
\centering
\caption{The dashed curve illustrates the deflection angle in the absence of plasma, see Ref.\ \cite{Bozza2002}. The solid curves represent the deflection angles in Schwarzschild spacetime in the presence of a non-uniform plasma for different values of $q$. In all cases we set $k = 0.1$. It is worth noting that the vacuum case and the case with $q=2$ both diverge at $r_0 = r_m = 3/2$.}
\end{figure}
\begin{align}
    a(q) &\simeq 2 + \frac{2^{q - 1}}{3^{q + 2}}\left(q^2 - 7q + 4\right)k, \label{a(q)} \\
    b_D (q) &\simeq a(q)\log2, \label{bD(q)}
\end{align}
respectively.

The crucial step is the calculation of $b_R$. Indeed, in this case the resulting integral is not analytically tractable. The strategy we adopt is to first expand the integrand in powers of $\epsilon$ and then evaluate the integral. We first recall that $b_R$ is defined as
\begin{equation}\label{brdef}
    b_R = \int_0^1 g(z,r_m)dz,
\end{equation}
where the function $g(z,r_m)$ is given by
\begin{equation}
    g(z,r_m) = 2n_m\left(f(z,r_m) - f_0 (z,r_m)\right).
\end{equation}
Explicitly writing down the expressions for $f(z,r_m)$ and $f_0 (z,r_m)$ (given by Eqs.\ \eqref{f} and \eqref{f_0}, respectively, specialized to this case), linearizing the integrand in \eqref{brdef} with respect to $\epsilon$ and setting $\epsilon = 1$, leads to
\begin{equation}
    b_{R}(q) \simeq b_{R,0} + b_{R,1}(q)k,
\end{equation}
with the second term in the above expression defined by
\begin{equation}\label{integral general case}
    b_{R,1}(q) \coloneqq \int_0^1 \mathrm{b}_{R,1}(z,q)dz,
\end{equation}
where, in turn, $\mathrm{b}_{R,1}(z,q)$ reads
\begin{multline}
    \mathrm{b}_{R,1}(z,q) \coloneqq \frac{2^q 3^{-\frac{3}{2} - q}}{\left[z^2\left(3 - 2z\right)\right]^{\frac{3}{2}}}\left\{(1 - z)^q (3 + 6z) \right.
    \\
    \left. \hspace{2.1cm} + \hspace{1mm} z(q - 2)\left[z(z - 3) + 3\right] \right\}
    \\
    -\hspace{0.3mm}\frac{2^{q - 1}}{3^{q + 2}}\frac{q^2 - 7q + 4}{z}.
\end{multline}
The integral in \eqref{integral general case} cannot be solved exactly for every $q > 0$. Nevertheless, by restricting our attention to values of $q$ in the interval $[0.5,5]$, we are able to find a good approximation of the integral in terms of polynomial functions, that is
\begin{multline}\label{br1h}
    b_{R,1}(q)\bigr\rvert_{q \in [0.5,5]} \simeq 2.60655 \times 10^{-6}
    \\
    \hspace{1.05cm}\times \left(q^2 - 10.8264q + 33.1271\right)
    \\
    \hspace{1.05cm}\times\left(q^2 - 6.60733q + 19.4469\right)
    \\
    \hspace{1.05cm} \times\left(q^2 - 12.5198q + 39.4461\right)
    \\
    \hspace{1.2cm}\times\left(q^2 - 0.699467q + 6.59669\right)
    \\
    \times\left(q - 2.56535\right)\left(q - 0.219367\right).
\end{multline}
For the values of $q$ that we considered in the previous sections, the above formula gives
\begin{figure}[t]
\includegraphics[width=8cm]{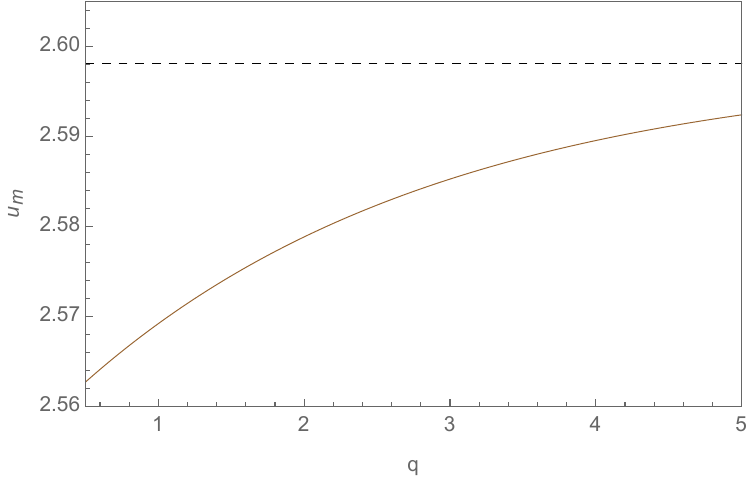}
\centering
\caption{Minimum impact parameter defined in Eq.\ \eqref{mip} as a function of $q$, with $k$ set to $0.1$. The dashed line represents the value in the absence of plasma.}
\end{figure}
\begin{align*}
    b_{R,1}(1) &\approx -0.20, \\
    b_{R,1}(2) &\approx -0.07, \\
    b_{R,1}(3) &\approx 0.04,
\end{align*}
thus finding perfect agreement, at least up to the second decimal place considered here.
It is important to remark that, even if Eq.\ \eqref{br1h} is only valid for $q \in [0.5,5]$, such interval can be easily extended. 

We can now finally write the deflection angle for any value of $q$ in the interval $[0.5,5]$ as
\begin{multline} \label{daqr0}
    \hat{\alpha}(r_0,q,k) \simeq -a(q,k)\log\delta(r_0 ,q,k)
    \\
    + b_{D}(q,k) + b_{R,0} + b_{R,1}(q)k - \pi,
\end{multline}
where $\delta(r_0 ,q,k)$ is given by
\begin{equation}
    \delta(r_0 ,q,k) = \frac{r_0}{r_m (q, k)} - 1.
\end{equation}
Moreover, the coefficient $a(q,k)$ and the radius of the photon sphere $r_m (q,k)$ are given by Eqs.\ \eqref{a(q)} and \eqref{rmq}, respectively, while $b_{D}(q,k), b_{R,0}$ and $b_{R,1}(q,k)$ are given by \eqref{bD(q)}, \eqref{vacuum b_R} and \eqref{br1h}, respectively.

Fig.\ 5 illustrates how the radius of the photon sphere is approached for different values of $q$.

Below we also write down the formula for the deflection angle as a function of $q$ and $u$, that is
\begin{equation}\label{general formula da}
    \hat{\alpha}(u,q,k) \simeq -\Bar{a}(q,k)\log\varepsilon(u,q,k) + \Bar{b}(q,k),
\end{equation}
where $\varepsilon(u,q,k)$, $\Bar{a}(q,k)$ and $u_m(q,k)$ are given by
\begin{align}
    \varepsilon(u,q,k) &= \frac{u}{u_m(q,k)} - 1, \\
    u_m(q,k) &= \frac{3^{\frac{1}{2} - q}}{2}\left(3^{q + 1} - 2^{q - 1}k\right), \label{mip} \\
    \Bar{a}(q,k) &= 1 + \frac{2^{q - 2}}{3^{q + 2}}\left(q^2 - 7q + 4\right)k, \label{bara}
\end{align}
respectively, while the coefficient $\Bar{b}(q,k)$ is
\begin{multline} \label{barb}
    \Bar{b}(q,k) = -\hspace{0.7mm}\pi + \log\left[216\left(7 - 4\sqrt{3}\right)\right]
    \\
    \hspace{0.5cm}+ \frac{2^{q - 2}}{3^{q + 2}}\left\{4\log6 - 16 + q\left[q\left(\log6 - 2\right)\right. \right.
    \\
    \left. \left.
    +\hspace{0.5mm}18 - 7\log6 \right] + b_{R,1}(q) \right\}k.
\end{multline}

We conclude the discussion of this section by showing the behavior of the coefficients $u_m ,\Bar{a}$ and $\Bar{b}$ as a function of $q$, setting $k = 0.1$ (Figs.\ 6, 7 and 8), and as a function of $k$ (Figs.\ 9, 10 and 11), respectively.

\begin{figure}[t]
\includegraphics[width=8cm]{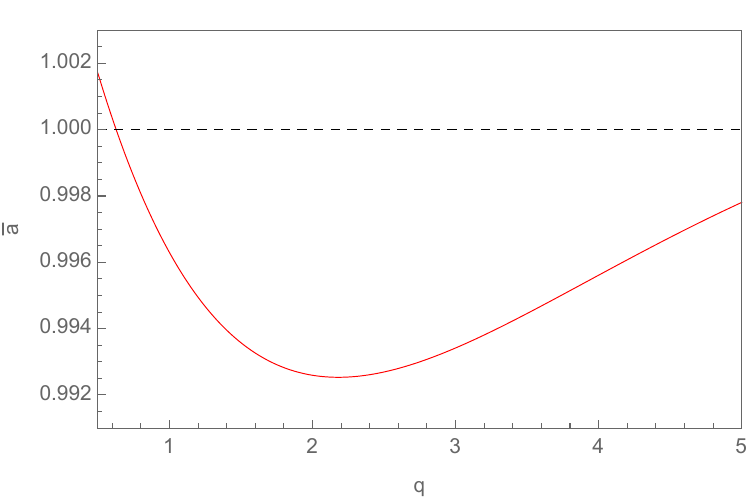}
\centering
\caption{Coefficient $\Bar{a}$ defined in Eq.\ \eqref{bara} as a function of $q$, with $k$ set to $0.1$. The dashed line represents the value in the absence of plasma.}
\end{figure}

\begin{figure}[t]
\includegraphics[width=8cm]{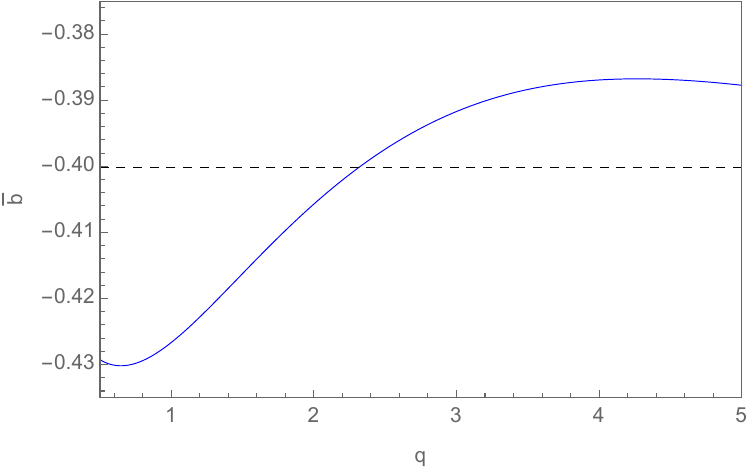}
\centering
\caption{Coefficient $\Bar{b}$ defined in Eq.\ \eqref{barb} as a function of $q$, with $k$ set to $0.1$. The dashed line represents the value in the absence of plasma.}
\end{figure}

\begin{figure}[t!]
\includegraphics[width=8cm]{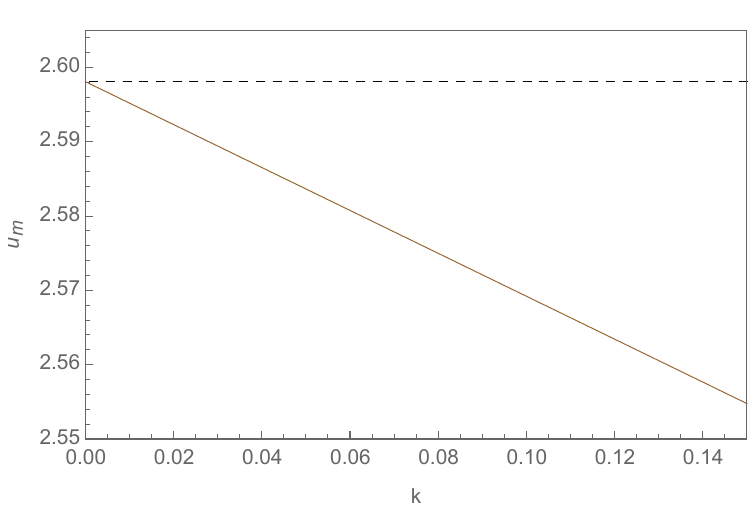}
\centering
\caption{Minimum impact parameter defined in Eq.\ \eqref{mip} as a function of $k$, setting $q = 1$. The dashed line represents the value in the absence of plasma.}
\end{figure}

\section{Observables} \label{sec:relimages}
In this section, we make use of the results from the previous section to calculate the properties of high-order images. Specifically, Eq.\ \eqref{general formula da} can be used to determine the positions and magnifications of such images when a Schwarzschild black hole is surrounded by inhomogeneous plasma.

As is well-known, gravitational lensing of a distant source by a black hole (`lens') leads to the formation of two infinite sequences of images on either side of the lens. These sequences comprise the primary image, the secondary image, and the high-order images. In cases of perfect alignment, an infinite sequence of Einstein rings occurs, including a `main' Einstein ring formed by merging the primary and secondary images, as well as higher-order rings. In the formation of high-order images, light orbits the black hole at least once, traveling very close to the photon sphere. The deflection angle for these images can be accurately calculated using the strong deflection limit \cite{Bozza2002, Bozza2010}.

Building on the results of Sec.\ \ref{subsec:nonuniformplh}, we incorporate the influence of plasma as a linear correction (low-density approximation), assuming that the number of images in the presence of plasma remains the same, but their position and magnification slightly change.

The general lens equation for spherically symmetric black holes can be written as \cite{BozzaScarpetta2007}
\begin{equation} \label{obs-01}
    \phi_O - \phi_S = \hat{\alpha}(u,q,k) + \pi \quad \text{mod} \hspace{1mm} 2\pi,
\end{equation}
where $\phi_O$ and $\phi_S$ are the azimuthal coordinates of the observer and the source, respectively. It is assumed that both the source and the observer are far away from the black hole ($D_{LS}, D_{OS} \gg 1$). Additionally, we remind that $u$ represents the impact parameter, Eq.\ \eqref{impactpar}, $q$ denotes the power-law index in the plasma distribution (see Sec.\ \ref{subsec:nonuniformplh}), and $k$ is the constant characterizing the plasma influence.

Fixing the origin of the azimuthal coordinate in such a way that $\phi_O = \pi$ and using Eq.\ \eqref{general formula da}, we can solve for the impact parameters of the high-order images, obtaining
\begin{equation} \label{obs-02}
    u_n (q,k) = u_m (q,k)\left(1 + l(q,k,n)\right),
\end{equation}
with the quantity $l(q,k,n)$ defined by
\begin{equation}
    l(q,k,n) \coloneqq \exp\left(\frac{\Bar{b}(q,k) + \phi_S - 2\pi n}{\Bar{a}(q,k)}\right).
\end{equation}
Above, $n$ denotes the number of loops around the black hole performed by light rays before reaching the observer and $\phi_S \in [-\pi, \pi]$. Eq.\ \eqref{obs-02} describes the images on one side of the lens; images on the other side of the lens can be found by substituting $\phi_O - \phi_S$ with $2\pi - \phi_O + \phi_S$. Therefore, every number $n \ge 0$ corresponds to a pair of images on different sides of the lens (see, e.g., pp.\ 2274--2275 of Ref.\ \cite{Bozza2010}). Eq.\ \eqref{obs-02} applies only to higher-order images ($n \ge 1$), while for primary and secondary images (a pair with $n=0$), the strong deflection limit approximation does not apply. For completeness, it should be noted that this widely used definition of $n$ should not be confused with the number of half-orbits, which has been recently used in describing higher-order photon rings in black hole images (see, e.g., \cite{Johnson-2020, Broderick-2022, BK-Tsupko-2022}).

Recalling that the angular separation of the image from the center of the lens is $\theta = u/D_{OL}$, where $D_{OL}$ is the distance between the lens and the observer (in our case, $D_{OL} \gg 1$), from the above equation we easily get
\begin{equation}\label{angular positions}
\theta_n (q,k)
= \frac{u_m (q,k)}{D_{OL}}\left(1 + l(q,k,n)\right).
\end{equation}

Computing the impact parameters in the presence of plasma for different values of $q$ in the range $[0.5,5]$, we find that they are always smaller than the corresponding impact parameters in vacuum; in Table I, considering specific values of $q$, some results for the impact parameters are presented. By varying $q$, we can therefore deduce that the presence of an inhomogeneous plasma reduces the angular size of the higher-order images. This conclusion is valid within the low-density plasma approximation. We also emphasize that here we restrict ourselves to considering a decreasing density profile, which is the most physically motivated situation. The decrease in angular size of higher-order images agrees with the results of Ref.\ \cite{Perlick2015}, where the decrease in angular size of the shadow in non-homogeneous plasma is found.

Let us conclude by considering the simple scenario where only the first ($n=1$) higher-order image is resolved in observations, while all others ($n\geq 2$) are grouped together near the shadow boundary. Recall that we consider only one side of the lens. The image with $n=1$ is the outermost among all higher-order images and has an angular separation $\theta_1$ from the center, as given by Eq.\ \eqref{angular positions}. All other images are closer to the shadow boundary, so we can approximate their asymptotic position as $\theta_{\infty} = u_m / D_{OL}$, where $\theta_{\infty}$ is the angular size of the shadow. This allows us to examine how the relative separation between the first image and the others, defined as
\begin{equation}\label{epsilon}
    s(q,k) \coloneqq l(q,k,1) = \exp\left(\frac{\Bar{b}(q,k) + \phi_S - 2\pi}{\Bar{a}(q,k)}\right),
\end{equation}
changes due to the presence of inhomogeneous plasma. Fig.\ 12 illustrates how the quantity $s(q,k)$ behaves as a function of $q$, where, for simplicity, we set $\phi_S = 0$ (in this case we thus have $\phi_O - \phi_S = \pi$, i.e., the background source, the lens and the observer are in perfect alignment).

Another important observable is the magnification of the higher-order images. For sources and observers very far from the black hole, it is given by \cite{Ohanian1987,Bozza2002,Bozza2010}
\begin{equation} \label{mu-n}
    \mu_n (q,k) = \left(\frac{D_{OS}}{D_{LS}}\right)^2 \frac{u_m^2 (q,k)s(q,k)}{D_{OL}^2 \Bar{a}(q,k)\sin\left(\pi - \phi_S \right)},
\end{equation}
with $D_{OS}$ being the distance between the observer and the source, $D_{LS}$ the one between the lens and the source, and finally $D_{OL}$ the one between the observer and the lens, already introduced before. Eq.\ \eqref{mu-n} was derived for the vacuum case, but it can also be applied to the plasma case if the quantities $u_m(q,k)$, $\bar{a}(q,k)$, and $\bar{b}(q,k)$ are now given by Eqs.\ \eqref{mip}, \eqref{bara} and \eqref{barb}, respectively. We also recall that we fixed the origin of the azimuthal coordinate in such a way that $\phi_O = \pi$. Setting $k = 0$ gives the expression for the magnification in the absence of plasma, that is
\begin{equation}
    \mu_{n,vac} = \left(\frac{D_{OS}}{D_{LS}}\right)^2 \frac{u_{m,vac}^2 \, s_{vac}}{D_{OL}^2 \sin\left(\pi - \phi_S \right)},
\end{equation}
where the subscript \textit{vac} stands for ``vacuum'' and the quantities $s_{vac}$ and $u_{m,vac}$ are given by \cite{Bozza2002}
\begin{equation}
    s_{vac} = 216(7 - 4\sqrt{3})e^{-\pi (2n + 1)}, \quad
    u_{m,vac} = \frac{3\sqrt{3}}{2},
\end{equation}
respectively.

In Table II, ratios $\mu_n / \mu_n^{vac}$ for different values of $q$ are presented. In Fig.\ 13, with $n = 1$ set, the ratio between the magnification factor for gravitational lensing in inhomogeneous plasma and that in vacuum is plotted as a function of $q$, concluding that the presence of inhomogeneous plasma reduces the magnifications of higher-order images, at least in the range of $q$ considered here.
\begin{figure}[t]
\includegraphics[width=8cm]{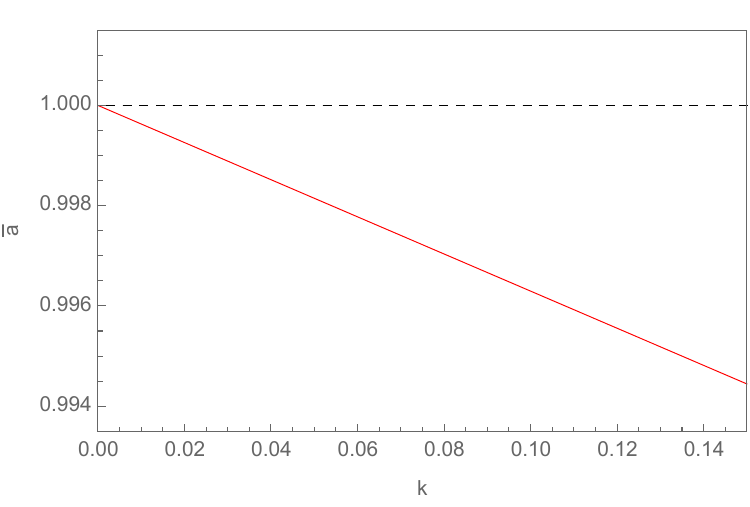}
\centering
\caption{Coefficient $\Bar{a}$ defined in Eq.\ \eqref{bara} as a function of $k$, setting $q = 1$. The dashed line represents the value in the absence of plasma.}
\end{figure}

\begin{figure}[t!]
\includegraphics[width=8cm]{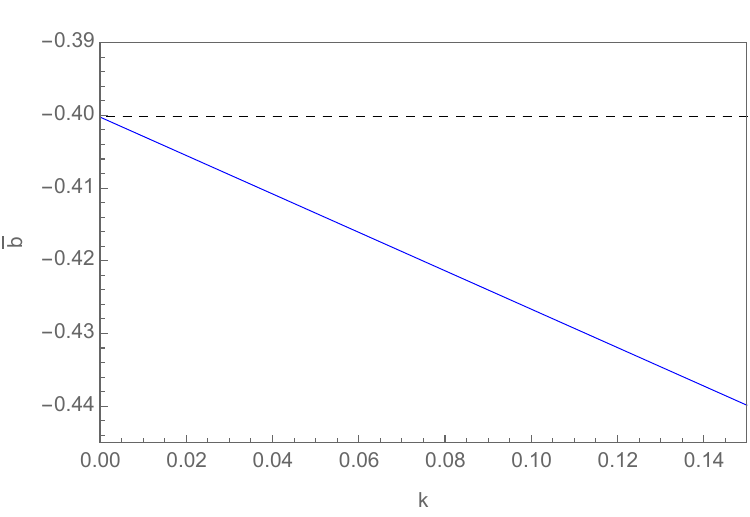}
\centering
\caption{Coefficient $\Bar{b}$ defined in Eq.\ \eqref{barb} as a function of $k$, setting $q = 1$. The dashed line represents the value in the absence of plasma.}
\end{figure}

\begin{figure}[t!]
\includegraphics[width=8cm]{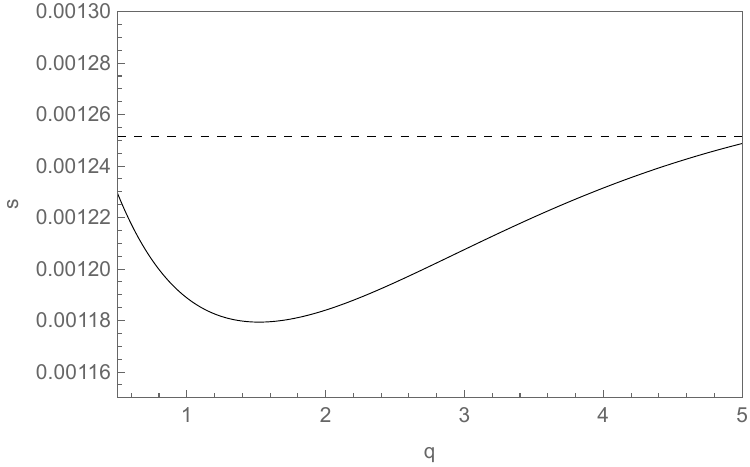}
\centering
\caption{Behavior of $s(q,k)$, defined in Eq.\ (\ref{epsilon}), is examined as a function of the power-law index $q$ with a fixed value of $k$. The constant $k$ is set to $0.1$, and the dashed line represents the function's value in the absence of plasma.}
\end{figure}

\begin{figure}[t!]
\includegraphics[width=8cm]{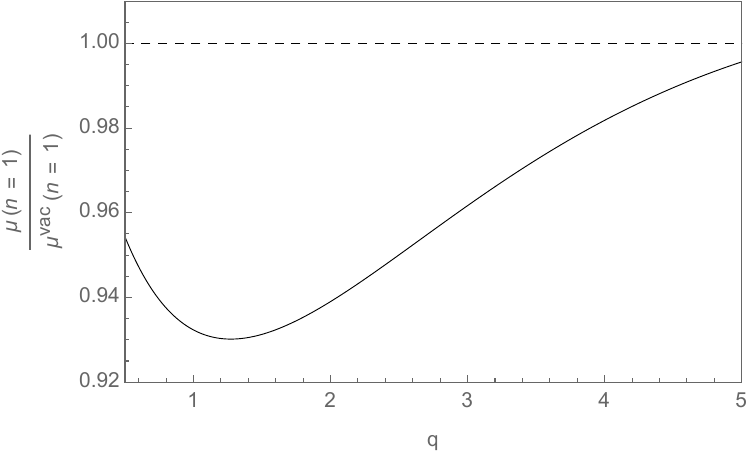}
\centering
\caption{Ratio between the magnification factors for lensing in the presence of inhomogeneous plasma and that in vacuum, as a function of $q$, with $n = 1$ and $k = 0.1$.}
\end{figure}

\section{Discussion and conclusions} \label{sec:concl}
This paper marks the initial stages of examining the deflection angle of photons by very compact objects in the presence of non-uniform plasma through analytical calculations. We find the deflection angle in the strong deflection limit, which corresponds to light rays circling several times around a compact object, and calculate properties of high-order images formed by such rays. In comparison with previous studies where the strong deflection limit was considered mainly in vacuum or in homogeneous plasma, here we develop a much more general approach valid for non-homogeneous plasma distribution.

We investigate analytically a quite general scenario: deflection angle in a static, asymptotically flat and spherically symmetric spacetime filled with non-homogeneous plasma having a spherically symmetric density profile (Sec.\ \ref{sec:sdlpl}), see Eq.\ \eqref{deflection angle} and Eq.\ \eqref{deflection angle imppar}. We focus on the Schwarzschild spacetime, providing insights into how different plasma density profiles impact the deflection of photons for this metric (Sec.\ \ref{sec:schwlensing}). The inclusion of results for an arbitrary power-law radial density profile (Sec.\ \ref{subsec:nonuniformplh}) adds versatility to the framework, making it ready for application to a broad range of astrophysical settings. The formulas presented in the paper are then applied to the calculation of the positions and magnifications of higher-order images (Sec.\ \ref{sec:relimages}).

It is worth noting the different level of approximation used in the various cases considered. As explained in the text, the two crucial steps of the strong deflection limit procedure are the calculation of the radius of the photon sphere and the coefficient $b_R$ (Eqs.\ \eqref{photon sphere equation} and \eqref{regular term general}, respectively). In the case of homogeneous plasma (Sec.\ \ref{subsec:hompl}), as well as in the cases of power-law indexes $q = 1$ and $q = 2$ (Secs.\ \ref{subsec:nonuniformpl1} and \ref{subsec:nonuniformpl2}, respectively), both the radius of the photon sphere and the coefficient $b_R$ have been found without relying on any approximation other than the one dictated by the strong deflection limit procedure. In terms of the distance of closest approach, $r_0$, the deflection angle in the homogeneous plasma case is given by Eq.~\eqref{deflection angle hp}, while the deflection angles in the cases $q = 1$ and $q = 2$ are given by Eqs.~\eqref{daq1} and \eqref{da2exact}, respectively. In the $q = 3$ case (Sec.\ \ref{subsec:nonuniformpl3}), the low-density plasma approximation was additionally used to find the radius of the photon sphere. However, the coefficient $b_R$ could still be determined through exact integration. The resulting deflection angle in this case is given by Eq.~\eqref{daq3}. As one might expect, for arbitrary $q$ in the range $[0.5, 5]$, the radius of the photon sphere can only be found in the low-density plasma approximation, and the coefficient $b_R$ can only be computed by linearizing the integrand (see Sec.\ \ref{subsec:nonuniformplh}). The result for the deflection angle is given by Eq.~\eqref{daqr0}. For the sake of completeness, we also provide references to the expressions for the deflection angle in terms of the impact parameter; in the homogeneous plasma case, it is given by Eq.~\eqref{daimp}, while in the cases $q = 1, q = 2$ and $q = 3$ it is given by Eqs.~\eqref{daimp1}, \eqref{daimp2} and \eqref{daimp3}, respectively. Finally, for arbitrary $q \in [0.5, 5]$, we have obtained Eq.~\eqref{general formula da}.

By opting for analytical methods, we provided a clear and direct understanding of the relation between photon deflection and non-uniform plasma environments, opening avenues for further exploration and application. Beyond delving into more complex plasma models, the next logical step in our study involves extending the analysis to address axially symmetric and stationary solutions to Einstein field equations.

\begin{table}[h!]
\centering
\begin{tabular}{||c c c c c||}
 \hline
 Impact parameter & Vacuum & $q = \frac{3}{2}$ & $q = 2$ & $q = 3$ \\ [0.5ex] 
 \hline\hline
 $u_1$ & 2.60133 & 2.57754 & 2.58188 & 2.58837 \\ 
 \hline
 $u_2$ & 2.59808 & 2.57451 & 2.57884 & 2.58525 \\
 [0.4ex] 
 \hline
\end{tabular}
\caption{Comparison between the impact parameters of the higher-order images in vacuum and inhomogeneous plasma for different values of $q$, in the specific case of perfect alignement. The constant $k$ has been set to $0.1$.}
\end{table}

\begin{table}[h!]
\centering
\begin{tabular}{||c c c c c||} 
 \hline
 $\mu_n /\mu_n^{vac}$& $\mu_n^{vac}$ & $q = \frac{3}{2}$ & $q = 2$ & $q = 3$ \\ [0.5ex] 
 \hline\hline
 $\mu_1 /\mu_1^{vac}$ & $0.716 \times 10^{-11}$ & 0.93 & 0.94 & 0.96 \\ 
 \hline
 $\mu_2 /\mu_2^{vac}$ & $0.134 \times 10^{-13}$ & 0.89 & 0.90 & 0.92 \\
 [0.6ex] 
 \hline
\end{tabular}
\caption{Here we compare the magnification factors of relativistic images for lensing in inhomogeneous plasma, across different values of $q$, with those in vacuum. We consider the ratios $\mu_n / \mu_{n}^{vac}$, which depend only
on $q$ and $k$ (the constant $k$ is set to 0.1). The values $\mu_n^{vac}$ have been taken from \cite{Tsupko2013}.}
\end{table}

\section*{ACKNOWLEDGMENTS} \label{sec:acknowledgements}
\noindent FF acknowledges support from the University of Salerno and the Istituto Nazionale di Fisica Nucleare through the Commissione Scientifica Nazionale 4 Iniziativa Specifica “Quantum Gravity in the SKY”.\ FF also thanks Francesco Grippa for reading the manuscript. The stay of OYuT at ZARM, Bremen University, is supported by a Humboldt Research Fellowship for experienced researchers from the Alexander von Humboldt Foundation; OYuT thanks Claus Lämmerzahl for the great hospitality. OYuT is grateful to Volker Perlick for useful discussions.


\begin{thebibliography}{99}

\bibitem{Walsh} D.\ Walsh, R.\ F.\ Carswell, and R.\ J.\ Weymann, 0957 + 561 A, B: twin quasistellar objects or gravitational lens?, Nature \textbf{279}, 381--384 (1979).

\bibitem{Soucail} G.\ Soucail, Y.\ Mellier, B.\ Fort, and J.\ P.\ Picat, A blue ring-like structure in the center of the A 370 cluster of galaxies, Astronomy \& Astrophysics \textbf{172}, L14--L16 (1987). 

\bibitem{Paczynski} B.\ Paczynski,
Gravitational microlensing by the Galactic halo,
Astrophysical Journal \textbf{304}, 1--5 (1986).

\bibitem{Falco1992} P.\ Schneider, J.\ Ehlers, and E.\ Falco, \textit{Gravitational Lenses} (Springer-Verlag, Berlin Heidelberg, 1992).

\bibitem{Narayan} R.\ Narayan and M.\ Bartelmann, \textit{Lectures on gravitational lensing}, arXiv:astro-ph/9606001v2.

\bibitem{Wambsganss1998} J.\ Wambsganss, Gravitational lensing in astronomy, Living Reviews in Relativity \textbf{1}, 12 (1998).

\bibitem{Schneider2001} M.\ Bartelmann and P.\ Schneider, Weak gravitational lensing, Physics Reports \textbf{340}, 291--472 (2001). 

\bibitem{Mollerach} S.\ Mollerach and E.\ Roulet,  \textit{Gravitational Lensing and Microlensing} (World Scientific, 2002).

\bibitem{Dodelson2017} S.\ Dodelson, \textit{Gravitational Lensing} (Cambridge University Press, 2017).

\bibitem{Meneghetti2021} M.\ Meneghetti, \textit{Introduction to Gravitational Lensing} (Springer International Publishing, 2021).

\bibitem{L1} Event Horizon Telescope Collaboration, K.\ Akiyama, A.\ Alberdi, W.\ Alef, K.\ Asada, R.\ Azulay, A.-K.\ Baczko, D.\ Ball, et al., First M87 Event Horizon Telescope results. I. The shadow of the supermassive black hole,
The Astrophysical Journal Letters \textbf{875}, L1 (2019).

\bibitem{L2} Event Horizon Telescope Collaboration, K.\ Akiyama, A.\ Alberdi, W.\ Alef, K.\ Asada, R.\ Azulay, A.-K.\ Baczko, D.\ Ball, et al., First M87 Event Horizon Telescope results. II. Array and instrumentation,
The Astrophysical Journal Letters \textbf{875}, L2 (2019).

\bibitem{L3} Event Horizon Telescope Collaboration, K.\ Akiyama, A.\ Alberdi, W.\ Alef, K.\ Asada, R.\ Azulay, A.-K.\ Baczko, D.\ Ball, et al., First M87 Event Horizon Telescope results. III. Data processing and calibration,
The Astrophysical Journal Letters \textbf{875}, L3 (2019).

\bibitem{L4} Event Horizon Telescope Collaboration, K.\ Akiyama, A.\ Alberdi, W.\ Alef, K.\ Asada, R.\ Azulay, A.-K.\ Baczko, D.\ Ball, et al., First M87 Event Horizon Telescope results. IV. Imaging the central supermassive black hole,
The Astrophysical Journal Letters \textbf{875}, L4 (2019).

\bibitem{L5} Event Horizon Telescope Collaboration, K.\ Akiyama, A.\ Alberdi, W.\ Alef, K.\ Asada, R.\ Azulay, A.-K.\ Baczko, D.\ Ball, et al., First M87 Event Horizon Telescope results. V. Physical origin of the asymmetric ring,
The Astrophysical Journal Letters \textbf{875}, L5 (2019).

\bibitem{L6} Event Horizon Telescope Collaboration, K.\ Akiyama, A.\ Alberdi, W.\ Alef, K.\ Asada, R.\ Azulay, A.-K.\ Baczko, D.\ Ball, et al., First M87 Event Horizon Telescope results. VI. The shadow and mass of the central black hole, The Astrophysical Journal Letters \textbf{875}, L6 (2019).

\bibitem{Kocherlakota-2021} 
P.\ Kocherlakota, L.\ Rezzolla, H.\ Falcke, C.\ M.\ Fromm,
M.\ Kramer, Y.\ Mizuno, A. Nathanail, et al.\ (Event Horizon Telescope Collaboration),
Constraints on black-hole charges with the 2017 EHT observations of M87*,
Physical Review D \textbf{103}, 104047 (2021).

\bibitem{L12} Event Horizon Telescope Collaboration, K.\ Akiyama, A.\ Alberdi, W.\ Alef, J.\ C.\ Algaba, R.\ Anantua, K.\ Asada, R.\ Azulay, U.\ Bach, A.-K.\ Baczko, D.\ Ball, et al.,
First Sagittarius A* Event Horizon Telescope results. I. The shadow of the supermassive black hole in the center of the Milky Way,
The Astrophysical Journal Letters \textbf{930}, L12 (2022).

\bibitem{L13} Event Horizon Telescope Collaboration, K.\ Akiyama, A.\ Alberdi, W.\ Alef, J.\ C.\ Algaba, R.\ Anantua, K.\ Asada, R.\ Azulay, et al., First Sagittarius A* Event Horizon Telescope results. II. EHT and multiwavelength observations, data processing, and calibration,
The Astrophysical Journal Letters \textbf{930}, L13 (2022).

\bibitem{L14} Event Horizon Telescope Collaboration, K.\ Akiyama, A.\ Alberdi, W.\ Alef, J.\ C.\ Algaba, R.\ Anantua, K.\ Asada, R.\ Azulay, et al., First Sagittarius A* Event Horizon Telescope results. III. Imaging of the Galactic center supermassive black hole,
The Astrophysical Journal Letters \textbf{930}, L14 (2022).

\bibitem{L15} Event Horizon Telescope Collaboration, K.\ Akiyama, A.\ Alberdi, W.\ Alef, J.\ C.\ Algaba, R.\ Anantua, K.\ Asada, et al., First Sagittarius A* Event Horizon Telescope results. IV. Variability, morphology, and black hole mass, The Astrophysical Journal Letters \textbf{930}, L15 (2022).

\bibitem{L16} Event Horizon Telescope Collaboration, K.\ Akiyama, A.\ Alberdi, W.\ Alef, J.\ C.\ Algaba, R.\ Anantua, K.\ Asada, R.\ Azulay, et al., First Sagittarius A* Event Horizon Telescope results. V. Testing astrophysical models of the Galactic center black hole, The Astrophysical Journal Letters \textbf{930}, L16 (2022).

\bibitem{L17} Event Horizon Telescope Collaboration, K.\ Akiyama, A.\ Alberdi, W.\ Alef, J.\ C.\ Algaba, R.\ Anantua, K.\ Asada, R.\ Azulay, et al., First Sagittarius A* Event Horizon Telescope results. VI. Testing the black hole metric, The Astrophysical Journal Letters \textbf{930}, L17 (2022).

\bibitem{Agol2000} H.\ Falcke, F.\ Melia, and E.\ Agol, Viewing the shadow of the black hole at the Galactic center, The Astrophysical Journal Letters \textbf{528}, L13 (2000).

\bibitem{Bambi2019} C.\ Bambi, K.\ Freese, S.\ Vagnozzi, and L.\ Visinelli, Testing the rotational nature of the supermassive object M87* from the circularity and size of its first image,
Physical Review D \textbf{100}, 044057 (2019).

\bibitem{Tamburini1} F.\ Tamburini, B.\ Thidé, and M.\ Della Valle, Measurement of the spin of the M87 black hole from its observed twisted light,
Monthly Notices of the Royal Astronomical Society: Letters \textbf{492}, 1, L22--L27 (2020).

\bibitem{Tamburini2} F.\ Tamburini, F.\ Feleppa, and B.\ Thidé, Twisted light, a new tool for general relativity and beyond — Revealing the properties of rotating black holes with the vorticity of light —, International Journal of Modern Physics D \textbf{30}, 14, 2142017 (2021).

\bibitem{Tamburini3} F.\ Tamburini, F.\ Feleppa, I.\ Licata, and B.\ Thidé, Kerr-spacetime geometric optics for vortex beams, Physical Review A \textbf{104}, 013718 (2021).

\bibitem{Darwin1959} C.\ Darwin, The gravity field of a particle, Proceedings of the Royal Society of London, Series A, Mathematical and Physical Sciences \textbf{249}, 180 (1959).

\bibitem{Atkinson1965} R.\ d’E.\ Atkinson, On light tracks near a very massive star, Astronomical Journal \textbf{70}, 517 (1965).

\bibitem{Misner1973} C.\ W.\ Misner, K.\ S.\ Thorne, and J.\ A.\ Wheeler, \textit{Gravitation} (W. H. Freeman, San Francisco, 1973).

\bibitem{Luminet1979} J.-P.\ Luminet, Image of a spherical black hole with thin accretion disk,
Astronomy and Astrophysics \textbf{75}, 228--235 (1979).

\bibitem{Ohanian1987} H.\ C.\ Ohanian, The black hole as a gravitational ‘‘lens’’,
American Journal of Physics \textbf{55}, 428--432 (1987).

\bibitem{Ellis2000} K.\ S.\ Virbhadra and G.\ F.\ R.\ Ellis, Schwarzschild black hole lensing, Physical Review D \textbf{62}, 084003 (2000).

\bibitem{Frittelli2000} S.\ Frittelli, T.\ P.\ Kling, and E.\ T.\ Newman, Spacetime perspective of Schwarzschild lensing, Physical Review D \textbf{61}, 064021 (2000).

\bibitem{Perlick2004} V.\ Perlick, Exact gravitational lens equation in spherically symmetric and static spacetimes, Physical Review D \textbf{69}, 064017 (2004).

\bibitem{Bozza2010} V.\ Bozza, Gravitational lensing by black holes,
General Relativity and Gravitation \textbf{42}, 2269--2300 (2010).

\bibitem{Bozza2001} V.\ Bozza, S.\ Capozziello, G.\ Iovane, and G.\ Scarpetta,
Strong field limit of black hole gravitational lensing,
General Relativity and Gravitation \textbf{33}, 1535--1548 (2001).

\bibitem{Bozza2002} V.\ Bozza, Gravitational lensing in the strong field limit, Physical Review D \textbf{66}, 103001 (2002).

\bibitem{Bozza2003} V.\ Bozza, Quasiequatorial gravitational lensing by spinning black holes in the strong field limit, Physical Review D \textbf{67}, 103006 (2003).

\bibitem{Claudel2001} C.-M.\ Claudel, K.\ S.\ Virbhadra, and G.\ F.\ R.\ Ellis, The geometry of photon surfaces, Journal of Mathematical Phyics \textbf{42}, 818--838 (2001).

\bibitem{Hasse2002} W.\ Hasse and V.\ Perlick, Gravitational lensing in spherically symmetric static spacetimes with centrifugal force reversal, General Relativity and Gravitation \textbf{34}, 415--433 (2002).

\bibitem{Perlick2004-review}
V.\ Perlick, Gravitational lensing from a spacetime perspective, Living Reviews in Relativity \textbf{7}, 9 (2004).

\bibitem{Iyer2007} S.\ V.\ Iyer and A.\ O.\ Petters, Light’s bending angle due to black holes: from the photon sphere to infinity, General Relativity and Gravitation \textbf{39}, 1563--1582 (2007).

\bibitem{Keeton2008} K.\ S.\ Virbhadra and C.\ R.\ Keeton, Time delay and magnification centroid due to gravitational lensing by black holes and naked singularities, Physical Review D \textbf{77}, 124014 (2008).

\bibitem{Tsupko2008} G.\ S.\ Bisnovatyi-Kogan and O.\ Yu.\ Tsupko, Strong gravitational lensing by Schwarzschild black holes, Astrophysics \textbf{51}, 99--111 (2008).

\bibitem{Majumdar2009} N.\ Mukherjee and A.\ S.\ Majumdar, Rotating brane-world black hole lensing in the strong deflection limit, Gravitation and Cosmology \textbf{15}, 263--272 (2009).

\bibitem{Tarasenko2010} A.\ Tarasenko, Reconstruction of a compact object motion in the vicinity of a black hole by its electromagnetic radiation, Physical Review D \textbf{81}, 123005 (2010).

\bibitem{Eiroa2011} E.\ F.\ Eiroa and C.\ M.\ Sendra, Gravitational lensing by a regular black hole, Classical and Quantum Gravity \textbf{28}, 085008 (2011).

\bibitem{Wei2012} S.-W.\ Wei, Yu.-X.\ Liu, C.-E.\ Fu, and K.\ Yang, Strong field limit analysis of gravitational lensing in Kerr-Taub-NUT spacetime, Journal of Cosmology and Astroparticle Physics, 10 (2012) 053.

\bibitem{Zhang2015} G.\ Li, Y.\ Zhang, L.\ Zhang, Z.\ Feng, and X.\ Zu, Strong gravitational lensing in the Einstein-Proca theory, International Journal of Theoretical Physics \textbf{54}, 1245--1252 (2015).

\bibitem{Alhamzawi2016} A.\ Alhamzawi and R.\ Alhamzawi, Gravitational lensing in the strong field limit by modified gravity, General Relativity and Gravitation \textbf{48}, 167 (2016).

\bibitem{Tsukamoto2016} N.\ Tsukamoto, Strong deflection limit analysis and gravitational lensing of an Ellis wormhole, Physical Review D \textbf{94}, 124001 (2016).

\bibitem{Aldi-Bozza-2017} 
G.\ F.\ Aldi and V.\ Bozza, Relativistic iron lines in accretion disks: The contribution of higher order images in the strong deflection limit, Journal of Cosmology and Astroparticle Physics 02 (2017) 033.

\bibitem{Dai2018} D.-C.~Dai, D.~Stojkovic, and G.~D.~Starkman, Strong lensing constraints on modified gravity models, Physical Review D \textbf{98} 124027 (2018).

\bibitem{Aratore2021} F.\ Aratore and V.\ Bozza, Decoding a black hole metric from the interferometric pattern of the relativistic images of a compact source, Journal of Cosmology and Astroparticle Physics 10 (2021) 054.

\bibitem{Kuang2022} X.-M.\ Kuang, Z.-Y. Tang, B.\ Wang, and A.\ Wang, Constraining a modified gravity theory in strong gravitational lensing and black hole shadow observations, Physical Review D \textbf{106}, 064012 (2022).

\bibitem{Aratore-Bozza-2024}
F.~Aratore and V.~Bozza,
Analytical perturbations of relativistic images in Kerr space-time, arXiv eprint, arXiv:2403.10169 (2024).

\bibitem{Gralla2019} S.\ E.\ Gralla, D.\ E.\ Holz, and R.\ M.\ Wald, Black hole shadows, photon rings, and lensing rings, Physical Review D \textbf{100}, 024018 (2019).

\bibitem{Johnson-2020} M.\ D.\ Johnson, A.\ Lupsasca, A.\ Strominger, G.\ N.\ Wong, S.\ Hadar, D.\ Kapec, R.\ Narayan, A.\ Chael, C.\ F.\ Gammie, P.\ Galison, et al., Universal interferometric signatures of a black hole’s photon ring, Science Advances \textbf{6}, eaaz1310 (2020).

\bibitem{Gralla2020} S.\ E.\ Gralla and A.\ Lupsasca, Observable shape of black hole photon rings, Physical Review D \textbf{102}, 124003 (2020).

\bibitem{Lupsasca2020} S.\ E.\ Gralla, A.\ Lupsasca, D.\ P.\ Marrone, The shape
of the black hole photon ring: A precise test of strong- field general relativity, Physical Review D \textbf{102}, 124004 (2020).

\bibitem{Gralla-Lupsasca-2020}
S.\ E.\ Gralla and A.\ Lupsasca,
Lensing by Kerr black holes,
Physical Review D \textbf{101}, 044031 (2020).

\bibitem{Wielgus-2021}
M.\ Wielgus, Photon rings of spherically symmetric black holes and robust tests of non-Kerr metrics, Physical Review D \textbf{104}, 124058 (2021).

\bibitem{Broderick-2022}
A.\ E.\ Broderick, P.\ Tiede, D.\ W.\ Pesce, and R.\ Gold,
Measuring spin from relative photon ring sizes,
The Astrophysical Journal \textbf{927}, 6 (2022).

\bibitem{Ayzenberg-2022}
D.\ Ayzenberg, Testing gravity with black hole shadow subrings,
Class. Quant. Grav. \textbf{39}, 105009 (2022). 

\bibitem{Guerrero-2022}
M.\ Guerrero, G.\ J.\ Olmo, D.\ Rubiera-Garcia, and D.~S.-C.~ G\'{o}mez, Multiring images of thin accretion disk of a regular naked compact object, Physical Review D \textbf{106}, 044070 (2022).

\bibitem{BK-Tsupko-2022}
G.\ S.\ Bisnovatyi-Kogan and O.\ Y.\ Tsupko,
Analytical study of higher-order ring images of the accretion disk around a black hole,
Physical Review D \textbf{105}, 064040 (2022).

\bibitem{Tsupko-2022}
O.\ Yu.\ Tsupko,
Shape of higher-order images of equatorial emission rings around a Schwarzschild black hole: Analytical description with polar curves, Physical Review D \textbf{106}, 064033 (2022).

\bibitem{Eichhorn-2023}
A.\ Eichhorn, A.\ Held, and P.-V. Johannsen,
Universal signatures of singularity-resolving physics in photon
rings of black holes and horizonless objects,
Journal of Cosmology and Astroparticle Physics 01 (2023) 043.

\bibitem{Broderick-Salehi-2023}
A.\ E.\ Broderick, K.\ Salehi, and B.\ Georgiev,
Shadow implications: What does measuring the photon ring imply for gravity?, The Astrophysical Journal \textbf{958}, 114 (2023).

\bibitem{Kocherlakota-2024-1}
P.\ Kocherlakota, L.\ Rezzolla, R.\ Roy, and M.\ Wielgus,
Prospects for future experimental tests of gravity with black hole imaging: Spherical symmetry, Physical Review D \textbf{109}, 064064 (2024).

\bibitem{Kocherlakota-2024-2}
P.\ Kocherlakota, L.\ Rezzolla, R.\ Roy, and M.\ Wielgus, Hotspots and Photon Rings in Schwarzschild Black Hole Spacetimes, arXiv e-print, arXiv:2403.08862 (2024).

\bibitem{Aratore-Tsupko-Perlick-2024}
F.\ Aratore, O.\ Yu.\ Tsupko, and V.\ Perlick,
Constraining spherically symmetric metrics by the gap between photon rings, Physical Review D (accepted, 2024), arXiv:2402.14733.

\bibitem{Tsupko2009} G.\ S.\ Bisnovatyi-Kogan and O.\ Yu.\ Tsupko, Gravitational radiospectrometer, Gravitation and Cosmology \textbf{15}, 20--27 (2009).

\bibitem{BK-Tsupko-2010}
G.\ S.\ Bisnovatyi-Kogan and O.\ Yu.\ Tsupko, Gravitational lensing in a non-uniform plasma, Monthly Notices of the Royal Astronomical Society \textbf{404}, 1790 (2010).

\bibitem{Morozova2013} V.\ S.\ Morozova, B.\ J.\ Ahmedov, A.\ A.\ Tursunov, Gravitational lensing by a rotating massive object in a plasma, Astrophysics and Space Science \textbf{346}, 513--520 (2013).

\bibitem{Er2014} X.\ Er and S.\ Mao, Effects of plasma on gravitational lensing, Monthly Notices of the Royal Astronomical Society \textbf{437}, 2180--2186 (2014).

\bibitem{Tsupko2015} G.\ S.\ Bisnovatyi-Kogan and O.\ Yu.\ Tsupko, Gravitational lensing in plasmic medium, Plasma Physics Report \textbf{41}, 562--581 (2015).

\bibitem{Gallo2018} G.\ Crisnejo and E.\ Gallo, Weak lensing in a plasma medium and gravitational deflection of massive particles using the Gauss-Bonnet theorem. A unified treatment, Physical Review D \textbf{97}, 124016 (2018).

\bibitem{Gallo20191} G.\ Crisnejo, E.\ Gallo, and K.\ Jusufi, Higher order corrections to deflection angle of massive particles and light rays in plasma media for stationary spacetimes using the Gauss-Bonnet theorem, Physical Review D \textbf{100}, 104045 (2019).

\bibitem{Gallo20192} G.\ Crisnejo, E.\ Gallo, and A.\ Rogers, Finite distance corrections to the light deflection in a gravitational field with a plasma medium, Physical Review D \textbf{99}, 124001 (2019).

\bibitem{Tsupko2020} O.\ Yu.\ Tsupko and G.\ S.\ Bisnovatyi-Kogan, Hills and holes in the microlensing light curve due to plasma environment around gravitational lens, Monthly Notices of the Royal Astronomical Society \textbf{491}, 5636--5649 (2020).

\bibitem{Sun2023} J.\ Sun, X.\ Er, O.\ Yu.\ Tsupko, Binary microlensing with plasma environment – star and planet, Monthly Notices of the Royal Astronomical Society \textbf{520}, 994--1004 (2023).

\bibitem{BK-Tsupko-2023-time-delay}
G.\ S.\ Bisnovatyi-Kogan and O.\ Yu.\ Tsupko, Time delay induced by plasma in strong lens systems, 
Monthly Notices of the Royal Astronomical Society \textbf{524}, 3060--3067 (2023).

\bibitem{Perlick-2000}
V.\ Perlick, \textit{Ray Optics, Fermat's Principle, and Applications to General Relativity} (Springer Berlin Heidelberg, 2000).

\bibitem{Tsupko2013}
O.\ Yu.\ Tsupko and G. S.\ Bisnovatyi-Kogan, Gravitational lensing in plasma: Relativistic images at homogeneous plasma, Physical Review D \textbf{87}, 124009 (2013).

\bibitem{Perlick2015} V.\ Perlick, O.\ Yu.\ Tsupko, and G.\ S.\ Bisnovatyi-Kogan, Influence of a plasma on the shadow of a spherically symmetric black hole, Physical Review D \textbf{92}, 104031 (2015).

\bibitem{Rogers2015} A.\ Rogers, Frequency-dependent effects of gravitational lensing within plasma, Monthly Notices of the Royal Astronomical Society \textbf{451}, 17--25 (2015).

\bibitem{Rogers20171} A.\ Rogers, Escape and trapping of low-frequency gravitationally lensed rays by compact objects within plasma, Monthly Notices of the Royal Astronomical Society \textbf{465}, 2151--2159 (2017).

\bibitem{Rogers20172} A.\ Rogers, Gravitational lensing of rays through the levitating atmospheres of compact objects, Universe \textbf{3}, 3 (2017).

\bibitem{Perlick2017} V.\ Perlick and O.\ Yu.\ Tsupko, Light propagation in a plasma on Kerr spacetime: Separation of the Hamilton-Jacobi equation and calculation of the shadow, Physical Review D \textbf{95}, 104003 (2017).

\bibitem{Huang2018} Y.\ Huang, Y.-P.\ Dong and D.-J.\ Liu, Revisiting the shadow of a black hole in the presence of a plasma, International Journal of Modern Physics, D \textbf{27}, 12, 1850114 (2018).

\bibitem{Yan2019} H.\ Yan, Influence of a plasma on the observational signature of a high-spin Kerr black hole, Physical Review D \textbf{99}, 084050 (2019).

\bibitem{Kimpson20191} T.\ Kimpson, K.\ Wu, and S.\ Zane,
Spatial dispersion of light rays propagating through a plasma in Kerr space-time,
Monthly Notices of the Royal Astronomical Society \textbf{484}, 2411--2419 (2019).

\bibitem{Kimpson20192} T.\ Kimpson, K.\ Wu, and S.\ Zane, 
Pulsar timing in extreme mass ratio binaries: A general relativistic approach, Monthly Notices of the Royal Astronomical Society \textbf{486}, 360--377 (2019).

\bibitem{Babar2020} G.\ Z.\ Babar, A.\ Z.\ Babar, and F.\ Atamurotov, 
Optical properties of Kerr–Newman spacetime in the presence of plasma, European Physical Journal C \textbf{80}, 761 (2020).

\bibitem{Tsupko2021} O.\ Yu.\ Tsupko, Deflection of light rays by a spherically symmetric black hole in a dispersive medium, Physical Review D \textbf{103}, 104019 (2021).

\bibitem{Chowdhuri2021} A.\ Chowdhuri and A.\ Bhattacharyya, Shadow analysis for rotating black holes in the presence of plasma for an expanding universe, Physical Review D \textbf{104}, 064039 (2021).

\bibitem{Eiroa2021} Javier Badía and Ernesto F.\ Eiroa, Shadow of axisymmetric, stationary, and asymptotically flat black holes in the presence of plasma, Physical Review D \textbf{104}, 084055 (2021).

\bibitem{Li2022} Q.\ Li, Y.\ Zhu, and T.\ Wang, Gravitational effect of plasma particles on the shadow of Schwarzschild black holes, European Physical Journal C \textbf{82}, 2 (2022).

\bibitem{Bezdekova2022} B.\ Bezděková, V.\ Perlick, and J.\ Bičák, Light propagation in a plasma on an axially symmetric and stationary spacetime: Separability of the Hamilton–Jacobi equation and shadow, Journal of Mathematical Physics \textbf{63}, 092501 (2022).

\bibitem{Zhang2023} Z.\ Zhang, H.~Yan, M.~Guo and B.~Chen, Shadows of Kerr black holes with a Gaussian-distributed plasma in the polar direction,
Physical Review D \textbf{107}, 024027 (2023).

\bibitem{Briozzo2023} G.\ Briozzo, E.\ Gallo, and T.\ Mädler, Shadows of rotating black holes in plasma environments with aberration effects, Physical Review D \textbf{107}, 124004 (2023).

\bibitem{Gallo2023} G.\ Briozzo and E.\ Gallo, Analytical expressions for pulse profile of neutron stars in plasma environments, European Physical Journal C \textbf{83}, 165 (2023).

\bibitem{Eiroa2023} J.\ Badía and E.\ F.\ Eiroa, Shadows of rotating Einstein-Maxwell-dilaton black holes surrounded by a plasma, Physical Review D \textbf{107}, 124028 (2023).

\bibitem{Bezdekova-2023}
B.\ Bezd\v{e}kov\'{a} and J.\ Bi\v{c}\'{a}k, Light deflection in plasma in the Hartle-Thorne metric and in other axisymmetric spacetimes with a quadrupole moment, Physical Review D \textbf{108}, 084043 (2023).

\bibitem{Bezdekova-2024}
B.\ Bezd\v{e}kov\'{a}, O.\ Yu.\ Tsupko, and C.\ Pfeifer,
Deflection of light rays in a moving medium around a spherically symmetric gravitating object,
Physical Review D (accepted, 2024), arXiv:2403.16842.

\bibitem{Perlick-Tsupko-2024} V.\ Perlick and O.\ Yu.\ Tsupko, Light propagation in a plasma on Kerr spacetime.\ II. Plasma imprint on photon orbits, Physical Review D \textbf{109}, 064063 (2024).

\bibitem{Synge1960} J.\ L.\ Synge, \textit{Relativity:\ The General Theory} (North-Holland Publishing Company, Amsterdam, 1960).

\bibitem{Muhleman-Johnston-1966}
D.\ O.\ Muhleman and I.\ D.\ Johnston, Radio propagation in the solar gravitational field,
Physical Review Letters \textbf{17}, 455 (1966).

\bibitem{Bicak-1975}
J.\ Bi\v{c}\'{a}k and P.\ Hadrava,
General-relativistic radiative transfer theory in refractive and dispersive media,
Astronomy \& Astrophysics \textbf{44}, 389 (1975).

\bibitem{Breuer-Ehlers-1980}
R.\ A.\ Breuer and J.\ Ehlers,
Propagation of high-frequency electromagnetic waves through a magnetized plasma in curved space-time.\ I,
Proceedings of the Royal Society of London. Series A, Mathematical and Physical Sciences \textbf{370}, 389 (1980).

\bibitem{Breuer-Ehlers-1981a}
R.\ A.\ Breuer and J.\ Ehlers, Propagation of high-frequency electromagnetic waves through a magnetized plasma in curved space-time.\ II. Application of the asymptotic approximation, Proceedings of the Royal Society of London. Series A, Mathematical and Physical Sciences \textbf{374}, 1756 (1981).

\bibitem{Breuer-Ehlers-1981b}
R.\ A.\ Breuer, J.\ Ehlers,
Propagation of electromagnetic waves through magnetized plasmas in arbitrary gravitational fields,
Astronomy \& Astrophysics \textbf{96}, 293 (1981).

\bibitem{Bliokh-Minakov-1989}
P.\ V.\ Bliokh and  A.\ A.\ Minakov, \textit{Gravitational Lenses} (in Russian), (Naukova Dumka, Kiev, 1989).

\bibitem{Kulsrud-Loeb-1992}
R.\ Kulsrud and A.\ Loeb,
Dynamics and gravitational interaction of waves in nonuniform media, Physical Review D \textbf{45}, 525 (1992).

\bibitem{Brod-Blandford-2003a}
A.\ Broderick and R.\ Blandford,
Covariant magnetoionic theory -- I. Ray propagation,
Monthly Notices of the Royal Astronomical Society \textbf{342}, 4, 1280--1290 (2003).

\bibitem{Brod-Blandford-2003b}
A.\ Broderick and R.\ Blandford,
Covariant magnetoionic theory -- II. Radiative transfer, 
Monthly Notices of the Royal Astronomical Society \textbf{349}, 3, 994--1008 (2004).

\bibitem{Tsupko2017} G.\ S.\ Bisnovatyi-Kogan, O.\ Yu.\ Tsupko, Gravitational lensing in presence of plasma: Strong lens systems, black hole lensing and shadow, Universe \textbf{3}, 57 (2017).

\bibitem{Cunha2018} P.\ V.\ P.\ Cunha and C.\ A.\ R.\ Herdeiro, Shadows and strong gravitational lensing: a brief review, General Relativity and Gravitation \textbf{50}, 42 (2018).

\bibitem{Perlick2022} V.\ Perlick and O.\ Yu.\ Tsupko, Calculating black hole shadows: Review of analytical studies, Physics Reports \textbf{947}, 1--39 (2022).

\bibitem{BozzaScarpetta2007} V.\ Bozza and G.\ Scarpetta, Strong deflection limit of black hole gravitational lensing with arbitrary source distances, Physical Review D \textbf{76}, 083008 (2007).

\end{thebibliography}
\end{document}